\documentclass[11pt,twoside]{article}
\usepackage[pdftex]{graphicx}
\usepackage{amsmath}
\usepackage{amssymb}
\usepackage{cite}
\usepackage{xcolor}

\begin{document}
\newcommand{\pst}{\hspace*{1.5em}}

\newcommand{\rigmark}{\em Journal of Russian Laser Research}
\newcommand{\lemark}{\em Volume 30, Number 5, 2009}

%\lhead[\fancyplain{\rigmark, {\em \lemark}}{\rigmark}]{\fancyplain{\rigmark, {\em \lemark}}{\lemark}}
%\chead{}\rhead[\fancyplain{}{\lemark}]{\fancyplain{}{\rigmark}}
%\plainfootrulewidth 0.4pt
\newcommand{\be}{\begin{equation}}
\newcommand{\ee}{\end{equation}}
\newcommand{\bm}{\boldmath}
\newcommand{\ds}{\displaystyle}
\newcommand{\bea}{\begin{eqnarray}}
\newcommand{\eea}{\end{eqnarray}}
\newcommand{\ba}{\begin{array}}
\newcommand{\ea}{\end{array}}
\newcommand{\arcsinh}{\mathop{\rm arcsinh}\nolimits}
\newcommand{\arctanh}{\mathop{\rm arctanh}\nolimits}
\newcommand{\bc}{\begin{center}}
\newcommand{\ec}{\end{center}}

\renewcommand{\thefootnote}{\fnsymbol{footnote}}

\thispagestyle{plain}

\label{sh}

%\lfoot[\fancyplain{\ \\[1mm] \thepage}{\ \\[1mm]\thepage}]{\fancyplain{}{}}

\begin{center} {\Large \bf
\begin{tabular}{c}
EVOLUTION AND ENTANGLEMENT OF GAUSSIAN STATES 
\\[-1mm]
IN THE PARAMETRIC AMPLIFIER
\end{tabular}
 } \end{center}

\bigskip

\bigskip

\begin{center} {\bf
Julio A. L\'opez-Sald\'ivar$^{1}$\footnote{Julio A. L\'opez-Sald\'ivar e-mail:julio.lopez@nucleares.unam.mx}, Armando Figueroa$^1$, Octavio Casta\~nos$^1$,  \\
Ram\'on L\'opez-Pe\~na$^1$ Margarita A. Man'ko$^{2}$, Vladimir I. Man'ko$^{2,3}$
}\end{center}

\medskip

\begin{center}
{\it
$^1$Instituto de Ciencias Nucleares, Universidad Nacional
Aut\'onoma de M\'exico, Apdo. Postal 70-543 M\'exico 04510 D.F. \\
\smallskip
$^2$ P. N. Lebedev Physical Institute, Leninskii Prospect, 53,
Moscow 119991, Russia \\
$^3$ Moscow Institute of Physics and Technology (State University)
   Dolgoprudnyi, Moscow Region 141700, Russia
}
\smallskip

\end{center}

\begin{abstract}\noindent
The linear time-dependent constants of motion of the parametric
amplifier are obtained and used to determine in the tomographic-probability representation
the evolution of a general two-mode Gaussian state. By means of the
discretization of the continuous variable density matrix, the von
Neumann and linear entropies are calculated to measure the
entanglement properties between the modes of the amplifier.  The
obtained results for the nonlocal correlations are compared with
those associated to a linear map of discretized symplectic Gaussian-state tomogram
onto a qubit tomogram.  This qubit portrait procedure is used to
establish Bell-type's inequalities, which provide a necessary
condition to determine the separability of quantum states, which can be evaluated through homodyne detection. Other no-signaling nonlocal correlations are defined through the portrait procedure for
noncomposite systems.
\end{abstract}

\medskip


\section{Introduction}
The linear time-dependent invariants of multidimensional quadratic
Hamiltonians in the position and momentum operators have been the
subject of many research
studies~\cite{trifonov,manko,libro-manko,suslov, castanos}. These
constants of motion are useful to determine the propagators of the
Hamiltonian systems and thus to study time-dependent problems in
quantum mechanics. It has been shown that these propagators can be
also obtained via the path integral formulation of quantum
mechanics~\cite{feynman}.

There are many specific systems of physical interest described by
quadratic Hamiltonians, such as the parametric
amplifier~\cite{rekdal,walls}, and others used in circuit
electrodynamics based on the Josephson junction
technique~\cite{zeilinger1,zeilinger2,zeilinger3}.

The tomographic probability representation introduced in 1996 is a
generalization of the optical tomography scheme~\cite{mancini}. In
this new formulation of quantum mechanics, the states are described
by measurable positive probabilities.

In 1935, EPR and Schr\"odinger defined the entanglement concept as a strange phenomena~\cite{einstein,Schrod35}. In the sixties, Bell~\cite{bell} and
Clauser--Horne--Shimony--Holt~(CHSH)~\cite{clauser}  established
that the predictions of quantum theory cannot be accounted by any
local theory. In a typical Bell experiment~\cite{brune}, it has
been established an inequality valid for separable states of a
bipartite system with the bound~2.  An experimental check of the
violation of the Bell--CHSH inequality with the bound~2 was first
performed in \cite{aspect}. Since the nineties the entanglement properties of a quantum system have been used as a resource to do tasks as quantum cryptography, quantum teleportation, and measurements in quantum computation.  These facts lead to the growing interest in the identification of the quantum correlations or in the non-local behaviour of the quantum physics~\cite{brune, guhne}. The bound~2 of the Bell--CHSH inequality can be violated for
entangled states of a composite system, which exhibit strong
correlations of the subsystems. The paradigmatic upper bound of the CHSH inequality
$2\sqrt2$ was proved by Cirelson~\cite{cirelson} for quantum
correlations in a bipartite system. On the other hand, formally there exists even an upper bound~4 discussed in
%Popescu and Rohlich
\cite{popescu}, which corresponds to superquantum correlations
in the systems modeling some properties of two-qubit states. A
tomographic approach to test the nonlocality was proposed for the
description of a correlated two-mode quantum state of
light~\cite{banaszek}. This proposal was implemented
in~\cite{bellini} using a balanced homodyne detection on temporal
modes of light, in which a clear violation of Bell's inequality was
found.

A qubit-portrait scheme of qudit tomograms has been proposed
in~\cite{chernega}, which allows one to discuss the Bell--CHSH
inequality~\cite{bell,clauser,nielsen} for two qubits within the
framework of the probability representation of quantum mechanics
introduced in~\cite{mancini,ibort}. It was also proposed that the
necessary condition of separability of a bipartite qudit state is
the separability of its qubit portrait~\cite{chernega}. This
portrait method can be extended to photon-number tomograms with some
modifications, which is useful to detect entanglement of two-mode
light states. Specifically, the violation of the Bell--CHSH
inequality indicates immediately that the state is
entangled~\cite{filippov}.

The models of the parametric amplifier and the frequency converter
were proposed in \cite{louisell} for two modes of the
electromagnetic field, which represent the signal and idler photons.
These harmonic oscillator modes are coupled by a classical field of
frequency $\omega$ (referred as pump frequency)  which may (or
may not) satisfy the resonance
condition~\cite{rekdal,mollow1,mollow2}, that is, $\omega=\omega_a +
\omega_b$ for the parametric amplifier or $  \omega= \omega_a -
\omega_b$ for the frequency converter.

The model Hamiltonian~\cite{rekdal} contains the main elements for
the description of physical realizations of the parametric
amplifier. As an example, we have a lossless nonlinear dielectric
substance that couples the modes of a resonant cavity with
reflecting walls~\cite{louisell}. In this case, there is a pump
field oscillating with a frequency equal to the sum of the
frequencies of the two modes, and it is strong enough to be
represented in classical terms. The two-mode nonresonant parametric
amplifier has been studied to find nonclassical features as revivals
and squeezing in~\cite{rekdal}, where it was shown that the
existence of quantum revivals is possible and the correlation
effects are very sensitive to the form of the initial state of the
system.

Recently, fiber optical parametric amplifiers have been developed
with a total amplification of 60--70~dB over an input
signal~\cite{marhic,marhic2}. Also the dynamics of the entanglement
of Gaussian states of systems in a reservoir model has been studied in
\cite{isar1,isar2}. The properties of superpositions of coherent
states are presented in \cite{calo95,julio}. A new resonant
condition technique has been developed to determine experimentally
the quadrature fluctuations of the light field, that is, the
covariance matrix~\cite{korolkova,marino,fabre}. Thus, the operation
of the parametric amplifier is sufficiently well understood.
Nevertheless, the dynamical invariants of this system were not
discussed in the literature, and one of the aims of this paper is to
obtain and apply the special linear-in-quadrature time-dependent
constants of motion of this amplifier.

%%%%%%%%%%%%%%%%%%%%%%%%%%%%
In this work, we construct the linear time-dependent constants of
motion of the nondegenerated parametric amplifier for the
trigonometric and hyperbolic cases. Using these constants of motion
we calculate, for the first time, the amplifier symplectic tomograms
associated to the dynamics of Gaussian states. The other goal of
this article is to propose a discretization method of the amplifier
density matrices, which allow us to evaluate the von Neumann and
linear entropies to measure the entanglement between the idler and
signal modes of the amplifier. A portrait map of the 
symplectic tomogram onto a qubit is defined to explore possible
sufficient conditions to have entanglement. For a noncomposite system, through the portrait picture, we show the possible existence of superquantum correlations i.e., the violation of the Cirelson bound.

The results presented could be of interest mainly in the context of theoretical and experimental
researches in the fields of the theory of entanglement and Bell non-locality.

This paper is organized as follows.

The Hamiltonian for the parametric amplifier is defined in
section~2, together with their analytic solution (trigonometric
case) in terms of the corresponding linear time-dependent
invariants. They are constructed for special conditions on the
Hamiltonian parameters, and the other solution (hyperbolic case) can
be obtained by analytic continuation. In section~3, the two-mode
Green function is written in terms of the time-dependent symplectic
matrix ($\boldsymbol{\Lambda}(t)$) and the corresponding two-mode Gaussian
states at time $t$ are presented. The evolution of the covariance
matrices are also calculated. The symplectic and optical tomograms
are described in section~4, making use of the covariance matrices.
In section~5, the von Neumann and linear entropies for the two-mode
Gaussian state in the parametric amplifier are calculated using the
discrete form of the two-mode density matrix and the corresponding
reduced density matrix for the subsystem. The portrait of continuous
symplectic and optical tomograms in the form of a two-qubit tomogram
is determined in section 6. This portrait is used to define Bell-type inequalities that is a sufficient condition for
separability, and a violation of this inequality is a necessary
condition to entanglement. Also in this section non-Bell
correlations within a noncomposite system are studied obtaining
strong correlations even larger that the Cirelson bound
$2\sqrt{2}$~\cite{cirelson}. In the final section, the conclusions
are given, and some technical details are presented in Appendix~A.

\section{Hamiltonian and linear dynamical invariants for the parametric amplifier}

The model of the parametric amplifier assumes that the two modes are
described by harmonic oscillators of frequencies $\omega_a$ and
$\omega_b$. The two modes are coupled by an oscillating parameter
called the pump with frequency $\omega$, which may (or may not)
satisfy the parametric resonance condition $\omega=\omega_a +
\omega_b$~\cite{mollow1, mollow2, rekdal}. Therefore, the Hamiltonian
can be written as
\begin{equation}
H=\hbar\omega_{a}a^{\dagger}a+\hbar\omega_{b}b^{\dagger}b-\hbar k
\left(a^{\dagger}b^{\dagger}e^{-i\omega t}+abe^{i\omega t}\right)\, ,
\label{1}
\end{equation}
where $k=\chi^{(2)} \sqrt{I_p}/v$, with $v$ being the group velocity
of the light in the medium, $\chi^{(2)}$ is the medium nonlinearity
and $I_p$ is the intensity of the pump~\cite{Levenson93}.  The sets
$(a^\dagger,a)$ and $(b^\dagger, b)$ define the photon creation and
annihilation operators for the two different electromagnetic modes.
The parameter $k$ is the coupling constant between the
electromagnetic modes. Notice that the operator $N_1 -N_2$ is a
constant of motion for the system.

The solution of the model Hamiltonian can be obtained by means of
several procedures: either by the use of the Heisenberg equations of
motion for the creation and annihilation operators, or by the use of
the interaction picture together with the Wei--Norman
procedure~\cite{louisell,mollow1, wei}. In this work, we use the
construction of the linear time-dependent invariants of the
system~\cite{manko}.

To obtain the linear in operators $a$, $a^\dagger$, $b$, $b^\dagger$
constants of motion for the parametric amplifier, one should solve a
system of differential equations given in Appendix~A; their solution
reads
\begin{eqnarray}
A(t) &=& e^{i \left(\Omega/2+\omega_{a} \right) t} \,
\left(\cos\nu t-\frac{i\Omega}{2\nu}\sin\nu t\right)a-
\frac{i \, k}{\nu}e^{-i \left(\Omega/2+\omega_{b} \right) t} \,
\sin\nu t \ b^{\dagger}, \nonumber \\
B(t) &=& e^{i \left(\Omega/2+\omega_{b} \right) t}
\left(\cos \nu t-\frac{i\Omega}{2\nu}\sin\nu t\right)b-
\frac{i \, k}{\nu}e^{-i \left(\Omega/2+\omega_{a} \right) t} \,
 \sin\nu t \ a^{\dagger} \, , \qquad
\label{2}
\end{eqnarray}
which, together with the corresponding creation operators
$A^\dagger$ and $B^\dagger$, satisfy the commutation relations of
boson operators $[A,A^{\dagger}]=1=[B,B^{\dagger}]$ with all others
commutators equal to zero. We define
$\Omega=\omega-\omega_{a}-\omega_{b}$, the detuning and
$\nu=\sqrt{\Omega^{2}/4-k^{2}}$ which we take as a real number. In the case of parametric
resonance, $\Omega=0$.  This system also has a solution for the
invariants where $\nu=\sqrt{k^2-\Omega^2/4}$ is real; in that case, the form
of the invariants can be obtained by analytic continuation, i.e.,
changing $\nu\rightarrow i \nu$ in Eq.~(\ref{2}) and substituting
trigonometric by hyperbolic functions, an analogous solution has
been obtained for Heisenberg operators in~\cite{walls}.

At multiples of the time $t= \pi /\nu$, the invariants take the form
\[
A( \pi n/\nu)=e^{i(\Omega/2+\omega_a)\pi n/\nu}(-1)^n \ a \, ,
\quad B( \pi n/\nu)=e^{i(\Omega/2+\omega_b)\pi n/\nu}(-1)^n \ b \, ,
\]
i.e., they are equal to the original boson operators multiplied by a
phase.

Establishing the relation between the creation and annihilation
invariant operators ($A^{\dagger}$, $B^{\dagger}$, $A$, $B$) with
the momentum and position quadrature operators ($P_1$, $P_2$, $Q_1$
and $Q_2$) representing the electric and magnetic fields in the
corresponding modes, one gets the matrix relation
\begin{equation}
\left(\begin{array}{c}
\mathbf{P}(t)\\
\mathbf{Q}(t)
\end{array}\right)=\mathbf{\Lambda}(t)\left(\begin{array}{c}
\mathbf{p}\\
\mathbf{q}
\end{array}\right) \equiv \left(\begin{array}{cc}
\boldsymbol{\lambda}_{1}(t) & \boldsymbol{\lambda}_{2}(t)\\
\boldsymbol{\lambda}_{3}(t) & \boldsymbol{\lambda}_{4}(t)
\end{array}\right) \, \left(\begin{array}{c}
\mathbf{p}\\
\mathbf{q}
\end{array}\right) \, ,
\label{3}
\end{equation}
where $\mathbf{P}(t)$, $\mathbf{Q}(t)$, $\mathbf{p}$, and
$\mathbf{q}$ denote $2 \times 1$ column vectors, while
$\mathbf{\Lambda}(t)$ defines a symplectic matrix in four dimensions
that satisfies the relation $\mathbf{\Lambda} (t) \mathbf{\Sigma}
\widetilde{\mathbf{\Lambda}} (t)=\mathbf{\Sigma}$, with the
definition
\[
\mathbf{\Sigma}=\left( \begin{array}{cc} 0 &\mathbf{I} \\
-\mathbf{I} & 0 \end{array} \right) \, ,
\]
where the $\mathbf{I}$ denotes a $2 \times 2$ identity matrix. Here,
$\mathbf{P}(t)$, $\mathbf{Q}(t)$ are the constants of motion with
initial conditions $\mathbf{P}(0)=\mathbf{p}$ and
$\mathbf{Q}(0)=\mathbf{q}$. For the parametric amplifier, the block
matrices for $\boldsymbol{\lambda}$ in units where $\hbar=\omega_a=1$, take the following form:
\begin{eqnarray}
\boldsymbol{\lambda}_1 =\left(\begin{array}{cc}
g_1(\omega_a) & g_2(\omega_b)/\sqrt{\omega_b} \\
g_2(\omega_a) \sqrt{\omega_b}& g_1 (\omega_b)
\end{array} \right) \ , \quad
\boldsymbol{\lambda}_2 =\left(\begin{array}{cc}
g_1(\omega_a +\frac{\pi}{2t}) & g_2(\omega_b-\frac{\pi}{2t})\sqrt{\omega_b} \\
g_2(\omega_a-\frac{\pi}{2t})\sqrt{\omega_b} & g_1 (\omega_b+\frac{\pi}{2t}) \omega_b
\end{array} \right) \ , \nonumber \\
\boldsymbol{\lambda}_3 =\left(\begin{array}{cc}
-g_1(\omega_a+\frac{\pi}{2t}) & g_2(\omega_b-\frac{\pi}{2t})/\sqrt{\omega_b} \\
g_2(\omega_a-\frac{\pi}{2t})/\sqrt{\omega_b} & -g_1 (\omega_b+\frac{\pi}{2t})/ \omega_b
\end{array} \right) \ , \quad
\boldsymbol{\lambda}_4 =\left(\begin{array}{cc}
g_1(\omega_a) & -g_2(\omega_b)\sqrt{\omega_b} \\
-g_2(\omega_a)/\sqrt{\omega_b} & g_1 (\omega_b)
\end{array} \right) \ ,
\label{lambdas}
\end{eqnarray}
with
\begin{eqnarray}
g_1 (\tilde{\omega})=\cos ((\Omega/2+\tilde{\omega}) t) \cos \nu t
+\frac{\Omega}{2 \nu}\sin ((\Omega/2+\tilde{\omega}) t) \sin \nu t \ , \nonumber \\
g_2 (\tilde{\omega})=\frac{k}{\nu} \sin ((\Omega/2+\tilde{\omega}) t) \sin \nu t \ . \nonumber
\end{eqnarray}

We point out that at the times $t=n\pi/\nu$, with $n$ an integer,
the invariants ($P$, $Q$) for modes $1$ and $2$ are determined by a
rotation of the original quadrature operators $p$ and $q$ of the
same mode;  the angles of these rotations are ($(\Omega
/2+\omega_a)n
\pi/\nu$) and ($(\Omega /2+\omega_b)n \pi/\nu$), respectively, i.e.,
\begin{eqnarray}
\left(\begin{array}{c}
P_j \\
Q_j
\end{array}\right)=
(-1)^n\left( \begin{array}{cc}
\cos ((\Omega /2+\omega_j)n \pi/\nu) & \omega_j \sin((\Omega /2+\omega_j)n \pi/\nu ) \\
-\sin ((\Omega /2+\omega_j)n \pi/\nu)/\omega_j & \cos((\Omega /2+\omega_j)n \pi/\nu)
\end{array} \right)
\left(\begin{array}{c}
p_j \\
q_j 
\end{array}\right) \ , \nonumber
\end{eqnarray}
with $j=1,\ 2$. These expressions are local transformations of the
original quadrature operators. Then we can expect that the
entanglement at these  times are equal to the entanglement at $t=0$,
independently of the initial state.

\section{Time evolution of two-mode Gaussian states}

In this section, the evolution of a two-dimensional Gaussian packet
in a parametric amplifier is studied. The time evolution of this
state is obtained by means of the Green function~\cite{manko}
\begin{equation}
 G(\mathbf{y},\,\mathbf{x};\,t)=\frac{i}{2\pi\sqrt{\det\boldsymbol{\lambda}_{3}}}
 \exp\left\{-\frac{i}{2}\Big(\widetilde{\mathbf{y}} \, \boldsymbol{\lambda}_{3}^{-1}
 \boldsymbol{\lambda}_{4} \,\mathbf{y}-2 \, \widetilde{\mathbf{y}} \,
 \boldsymbol{\lambda}_{3}^{-1} \, \mathbf{x}+\widetilde{\mathbf{x}} \,
 \boldsymbol{\lambda}_{1} \, \boldsymbol{\lambda}_{3}^{-1} \,\mathbf{x}\Big)\right\} \, ,
\label{8}
\end{equation}
where $\mathbf{y}$ and $\mathbf{x}$ are column vectors, the
$\widetilde{\mathbf{x}}$ means the matrix transposition of
$\mathbf{x}$, and the matrices $\boldsymbol{\lambda}_k$ were defined
in the previous section. The Green function of the amplifier is also
a Gaussian function of the two coordinates in the system, implying
that the evolution of a Gaussian state will be also a Gaussian
state.

The general two-mode Gaussian state is defined by \cite{manko}
\begin{equation}
\psi (\mathbf{x})=N \exp\left( -\widetilde{\mathbf{x}} \mathbf{A}_G
\mathbf{x}+\widetilde{\mathbf{B}}_G \mathbf{x}\right) \ ,
\label{9}
\end{equation}
with the normalization constant, $N=\sqrt{2/\pi}\
(\det{\mathbf{A}_G})^{1/4}\ e^{-\frac{1}{16} \,
\widetilde{\mathbf{B}}_G \, \mathbf{A}_G^{-1} \, \mathbf{B}_G}$;
$\widetilde{\mathbf{x}}=(x_{1},\, x_{2})$  is the transpose vector
of $\mathbf{x}$, $\widetilde{\mathbf{B}}_G=\left(B_1,\ B_2 \right)$,
and we took the matrix $\mathbf{A}_{G}$ as real and symmetric,
\begin{equation}
\mathbf{A}_G=\frac{1}{4}\left(\begin{array}{cc}
a_{11} & -a_{12}\\
-a_{12} & a_{22}
\end{array}\right).
\label{10}
\end{equation}

Using the propagator in Eq.~(\ref{8}), the time evolution
for the wave function is calculated in the integral form
\[
\psi(\mathbf{y},t)=\int d\mathbf{x}\,G(\mathbf{y},\mathbf{x};t)\,
\psi(\mathbf{x},0)\, ,
\]
giving the result for any quadratic Hamiltonian in the quadrature
components of the electromagnetic field:
\begin{equation}
\psi (\mathbf{y};t)=\frac{i N e^{-\frac{i}{2}\widetilde{\mathbf{y}}
\boldsymbol{\lambda}_3^{-1} \boldsymbol{\lambda}_4 \mathbf{y}}
e^{\frac{1}{4}(\widetilde{\mathbf{B}}_G+i \widetilde{\mathbf{y}}
\widetilde{\boldsymbol{\lambda}}_3^{-1})\left(\mathbf{A}_G+
\frac{i}{2}\boldsymbol{\lambda}_1 \boldsymbol{\lambda}_3^{-1}\right)^{-1}
(\mathbf{B}_G + i \boldsymbol{\lambda}_3^{-1}\mathbf{y})}}
{2\sqrt{\det (\boldsymbol{\lambda}_3) \det (\mathbf{A}_G+\frac{i}{2}
\boldsymbol{\lambda}_1 \boldsymbol{\lambda}_3^{-1})}} \ ;
\label{11}
\end{equation}
this expression will be used to calculate the density matrix and the
reduced density matrix for one mode in order to calculate the von
Neumann and linear entropies.

\subsection{Covariance matrix}

The evolution of the Gaussian in the parametric amplifier (even for
any quadratic Hamiltonian) is also a Gaussian state as seen in the
previous section. Any Gaussian state is completely determined by the
covariance matrix and the mean values of the position. The
covariance matrix of the general Gaussian state is calculated using
the constants of motion. The covariances and dispersions between the
quadrature components in the two-mode system can be put in the
matrix form as
\begin{eqnarray}
\boldsymbol{\sigma}_{pp}(t)&=&\left(\begin{array}{cc}
\sigma_{p_{1} p_{1}}(t) & \sigma_{p_{1} p_{2}}(t)\\
\sigma_{p_{2} p_{1}}(t) & \sigma_{p_{2} p_{2}}(t)
\end{array}\right),\:\boldsymbol{\sigma}_{qq}(t)=\left(\begin{array}{cc}
\sigma_{q_{1} q_{1}}(t) & \sigma_{q_{1} q_{2}}(t)\\
\sigma_{q_{2} q_{1}}(t) & \sigma_{q_{2} q_{2}}(t)
\end{array}\right),\nonumber\\ \boldsymbol{\sigma}_{pq}(t)&=&\left(\begin{array}{cc}
\sigma_{p_{1} q_{1}}(t) & \sigma_{p_{1} q_{2}}(t) \\
\sigma_{p_{2} q_{1}}(t) & \sigma_{p_{2} q_{2}}(t)
\end{array}\right) \, , \label{13}
\end{eqnarray}
with the usual definition $\sigma_{x,y}=\frac{1}{2}\langle\{
x,y\}\rangle-\langle x \rangle \langle y \rangle$, in terms of the
anti-commutator $\{ , \}$ of two operators.

The covariance matrix at $t=0$ can be defined by
\begin{equation}
\boldsymbol{\sigma}(0)=\left(\begin{array}{cc}
\boldsymbol{\sigma}_{pp0} & \boldsymbol{\sigma}_{pq0}\\
\widetilde{\boldsymbol{\sigma}}_{pq0} & \boldsymbol{\sigma}_{qq0}
\end{array}\right) \, , \label{14}
\end{equation}
with $\boldsymbol{\sigma}_{pp0}=\boldsymbol{\sigma}_{pp}(0)$,
$\boldsymbol{\sigma}_{qq0}=\boldsymbol{\sigma}_{qq}(0)$,
$\boldsymbol{\sigma}_{pq0}=\boldsymbol{\sigma}_{pq}(0)$ and
$\widetilde{\boldsymbol{\sigma}}_{pq}$ denoting the transpose of
$\boldsymbol{\sigma}_{pq}$; they satisfy $\sigma_{q_{1} q_{2}} =
\sigma_{q_{2} q_{1}}$, $\sigma_{q_{i} p_{j}} = \sigma_{p_{j} q_{i}}$
with $i,j =1,2$, and $\sigma_{p_{1}p_{2}} =\sigma_{p_{2}p_{1}}$.

By means of the expressions of the quadrature components in terms of
the linear time-dependent invariants, i.e., the inverse of
equation~(\ref{3}), it is straightforward to evaluate the covariance
matrix at time $t$ by the expression
\begin{equation}
\boldsymbol{\sigma}(t)=\mathbf{\Lambda}^{-1}\boldsymbol{\sigma}(0)
\widetilde{\mathbf{\Lambda}}^{-1} \, ,
\label{15}
\end{equation}
notice that $\det \boldsymbol{\sigma} (t)= \det \boldsymbol{\sigma} (0)=1/16$. The inverse
of the symplectic matrix $\mathbf{\Lambda}$ is given by
\begin{equation}
\mathbf{\Lambda}^{-1}=\left(\begin{array}{cc}
\widetilde{\boldsymbol{\lambda}}_{4} & -\widetilde{\boldsymbol{\lambda}}_{2}\\
-\widetilde{\boldsymbol{\lambda}}_{3} & \widetilde{\boldsymbol{\lambda}}_{1}
\end{array}\right)\,.
\label{16}
\end{equation}
Thus, the covariance matrix at time $t$ can be calculated for any
quadratic Hamiltonian and takes the form
\begin{eqnarray}
\boldsymbol{\sigma}_{pp}(t)=(\widetilde{\boldsymbol{\lambda}}_4
\boldsymbol{\sigma}_{pp0}-\widetilde{\boldsymbol{\lambda}}_2
\boldsymbol{\sigma}_{pq0})\boldsymbol{\lambda}_4+(-\widetilde{\boldsymbol{\lambda}}_4
\boldsymbol{\sigma}_{pq0}+\widetilde{\boldsymbol{\lambda}}_2
\boldsymbol{\sigma}_{qq0})\boldsymbol{\lambda}_2, \nonumber \\
\boldsymbol{\sigma}_{pq}(t)=(\widetilde{\boldsymbol{\lambda}}_4
\boldsymbol{\sigma}_{pp0}-\widetilde{\boldsymbol{\lambda}}_2
\boldsymbol{\sigma}_{pq0})\boldsymbol{\lambda}_3+(-\widetilde{\boldsymbol{\lambda}}_4
\boldsymbol{\sigma}_{pq0}+\widetilde{\boldsymbol{\lambda}}_2
\boldsymbol{\sigma}_{qq0})\boldsymbol{\lambda}_1, \nonumber \\
\boldsymbol{\sigma}_{qq}(t)=(\widetilde{\boldsymbol{\lambda}}_3
\boldsymbol{\sigma}_{pp0}-\widetilde{\boldsymbol{\lambda}}_1
\boldsymbol{\sigma}_{pq0})\boldsymbol{\lambda}_3+(-\widetilde{\boldsymbol{\lambda}}_3
\boldsymbol{\sigma}_{pq0}+\widetilde{\boldsymbol{\lambda}}_1
\boldsymbol{\sigma}_{qq0})\boldsymbol{\lambda}_1,
\label{17}
\end{eqnarray}
\noindent
where, for simplicity of the notation, we do not express the time
dependence in the matrices $\boldsymbol{\lambda}_k$. The covariance
matrix at time $t=0$ is given by the 2$\times$2 matrices
$\boldsymbol{\sigma}_{pp0}$, $\boldsymbol{\sigma}_{qq0}$ and
$\boldsymbol{\sigma}_{pq0}$. The covariance matrix at time $t$ is
calculated through Eq.~(\ref{17}) together with Eq.~(\ref{lambdas}).

Some examples of Gaussian states are given: One can obtain a
two-mode initial squeezed vacuum state making the substitution
\begin{equation}
\mathbf{A}_G=\frac{1}{2}\left( \begin{array}{cc}
\cosh 2 \, r & \sinh 2 \, r \sqrt{\omega_b} \\ \sinh 2 \, r \sqrt{\omega_b} & \cosh 2 \, r \, \omega_b \end{array} \right) ,
\quad \mathbf{B}_G=0 \  ,
\label{sqz-gauss}
\end{equation}
where $r\in \mathbb{R}$ is called the squeezing parameter. This
state is the result of the application of the two-mode squeeze
operator $S(r)=\exp{(r(ab-a^{\dagger}b^{\dagger}))}$ to the vacuum
state $\vert 0,0 \rangle$. This state can be written explicitly as
$\vert \beta \rangle=\sqrt{1-\beta^2}\sum_{n=0}^{\infty} \beta^{n}
\vert n, n\rangle$ with $\beta=-\textrm{tanh} r$. For this evolving
initial state, the block matrices for the covariance matrix at time
$t$ can be calculated using formula~(\ref{17}), obtaining
\begin{eqnarray}
\boldsymbol{\sigma}_{pp}(t)=\frac{1}{2(1-\vert \eta (t)\vert^2)}
\left( \begin{array}{cc} 1+\vert \eta (t)\vert^2 & -(\eta (t) +\eta^* (t)) \sqrt{\omega_b} \\
-(\eta (t) +\eta^* (t)) \sqrt{\omega_b} & (1+\vert \eta (t)\vert^2) \omega_b \end{array} \right) \ , \nonumber \\
\boldsymbol{\sigma}_{qq}(t)=\frac{1}{2(1-\vert \eta (t)\vert^2)}
\left( \begin{array}{cc} 1+\vert \eta (t)\vert^2 & (\eta (t) +\eta^* (t))/\sqrt{\omega_b} \\
(\eta (t) +\eta^* (t))/\sqrt{\omega_b} & (1+\vert \eta (t)\vert^2)/ \omega_b \end{array} \right) \ , \nonumber \\
\boldsymbol{\sigma}_{pq}(t)=\frac{1}{2(1-\vert \eta (t)\vert^2)}
\left( \begin{array}{cc} 0 & i (\eta^* (t) -\eta (t))/\sqrt{\omega_b} \\
i (\eta^* (t) -\eta (t))\sqrt{\omega_b} & 0 \end{array} \right) \ .
\label{20}
\end{eqnarray}
This covariance matrix corresponds to the squeezed vacuum state:
$\vert \eta (t) \rangle=\sqrt{1-\vert \eta (t)
\vert^2}\sum_{n=0}^{\infty} \eta^{n} (t) \vert n, n \rangle$ with
the squeeze parameter $\eta(t)$ given by
%\begin{equation}
%\eta (t)=\frac{i a_+ \cosh r-(e^{2 i a_z}+a_+ a_-)e^{i \phi}\sinh r}{\cosh r+i a_- e^{i \phi}\sinh r} e^{-i (\omega_a+\omega_b)t} \, .
%\label{21}
%\end{equation}
\begin{equation}
\eta (t)=\frac{e^{-i\omega t}}{2 k} \left( \frac{4 k^2 \exp \{ -2 \ln(\cos \nu t-
i\sin \nu t \tanh \gamma)\}}{\Omega -2 k \coth r+ 2 i \nu \tan (\nu
t+i\gamma)}- 2 i \nu \tan (\nu t +i \gamma)-\Omega\right) \ ,
\label{21}
\end{equation}
with $\gamma=\textrm{arctanh} (\Omega / 2 \nu)$. One can check that
for $t=0$, $\eta (0)=\beta=-\tanh r $.

The two-mode coherent state can be expressed by taking
\begin{equation}
\mathbf{A}_G=\frac{1}{2}\ \left(\begin{array}{cc}1 & 0 \\ 0 & \omega_b \end{array}\right), \quad
\mathbf{B}_G=\sqrt{2} \ ( \alpha_1, \sqrt{\omega_b}\, \alpha_2).
\label{coh-gauss}
\end{equation}
with $\alpha_1$ and $\alpha_2$ being the complex parameters for each
mode. The initial coherent state can be obtained in terms of the
translation operator $D(\alpha_1,\alpha_2)=\exp{(\alpha_1
a^{\dagger}-\alpha^*_1 a)}\exp{(\alpha_2 b^{\dagger}-\alpha^*_2 b)}$
applied to the vacuum state. The evolving coherent state has the
following covariance matrix
\begin{eqnarray}
\boldsymbol{\sigma}_{pp}(t)&=&\left(\begin{array}{cc}
\frac{\Omega^2-4 k^2 \cos2\nu t}{8 \nu^2} & f_1 (t) \sqrt{\omega_b}\\
f_1 (t) \sqrt{\omega_b}& \left( \frac{\Omega^2-4 k^2 \cos2\nu t}{8 \nu^2} \right)\omega_b
\end{array}\right) \, , \quad
\boldsymbol{\sigma}_{qq}(t)=\left(\begin{array}{cc}
\frac{\Omega^2-4 k^2 \cos2\nu t}{8 \nu^2} & -f_1 (t)/\sqrt{\omega_b}  \\
-f_1 (t)/\sqrt{\omega_b} &  \left(\frac{\Omega^2-4 k^2 \cos2\nu t}{8 \nu^2}\right)/ \omega_b
\end{array}\right) \, , \nonumber \\
 \boldsymbol{\sigma}_{pq}(t)&=&\left(
\begin{array}{cc} 0 & f_2 (t)/\sqrt{\omega_b} \\ f_2 (t)\sqrt{\omega_b} & 0 \end{array} \right) \ ,
\label{18}
\end{eqnarray}
with
\begin{eqnarray}
f_1 (t)=\frac{k}{2\nu^2}\left( \Omega \cos \omega t \sin^2 \nu t-
\nu \sin \omega t \sin 2 \nu t \right) \, , \nonumber\\
 f_2 (t)=\frac{k}{2\nu^2} \left( \Omega \sin \omega t \sin^2 \nu t +
  \nu \cos \omega t \sin 2 \nu t \right) \ ,
\label{19}
\end{eqnarray}
the covariance matrix of the system also can be used to study the
so-called tomographic representation of a Gaussian state determined
by the symplectic or optical tomogram. One can note that the covariance matrix for the coherent state is a periodic function, with period $T=\pi / \nu$ while this behavior is not present in the squeezed state.

%the covariance matrix of the two-mode squeezed vacuum state and the coherent state are used to define the symplectic and optical tomograms for both systems

\section{Symplectic and optical tomograms}

The two-mode symplectic tomographic distribution describes the
probability in the quadratures of the system in a rotated and
rescaled reference frame $(X_1, X_2)$ of the original quadratures
$(q_1,q_2,p_1,p_2)$, with the definition
\[
X_1=s_1q_1 \cos \theta_1 +s_1^{-1} p_1 \sin \theta_1,\quad
X_2=s_2q_2\cos\theta_2 +s_2^{-1}p_2\sin \theta_2 \ .
\]

According to \cite{mancini,ibort}, the symplectic tomographic
probability distribution can be determined by the expression
\begin{eqnarray}
& \mathcal{W}(X_{1},\mu_{1},\nu_{1};X_{2},\mu_{2},\nu_{2};
t) =\frac{1}{4\pi^{2}\left|\nu_{1}\nu_{2}\right|} \nonumber\\
& \times  \Big\vert\int\psi(y_{1},y_{2}; t)
\exp\Big(\frac{i\mu_{1}}{2\nu_{1}}y_{1}^{2}
+\frac{i\mu_{2}}{2\nu_{2}}y_{2}^{2}-\frac{iX_{1}y_{1}}{\nu_{1}}
-\frac{iX_{2}\,y_{2}}{\nu_{2}}\Big)\,dy_{1}\,dy_{2}\Big\vert^{2} \ ,
\label{29}
\end{eqnarray}
where $\mu_i=s_i \cos \theta_i$, $\nu_i=s_i^{-1} \sin \theta_i$.
This distribution, called the symplectic tomogram of two-mode system
state, is nonnegative and normalized, i.e.,
\[
\int\mathcal{W}(X_{1},\mu_{1},\nu_{1};X_{2},\mu_{2},\nu_{2}; t)dX_{1}dX_{2}=1.
\]

When a pure state is non-entangled, the tomogram can be expressed as
the multiplication of the distributions for each one of the modes
\cite{ibort}
\begin{equation}
\mathcal{W}(X_{1},\mu_{1},\nu_{1};X_{2},\mu_{2},\nu_{2}; t)=
\mathcal{W}_1(X_{1},\mu_{1},\nu_{1}; t) \mathcal{W}_2(X_{2},\mu_{2},\nu_{2}; t) \ ,
\label{separable}
\end{equation}
when $\mathcal{W}_1$ and $\mathcal{W}_2$ are the partial (also called reduced)
tomograms for the modes one and two and are defined as
\begin{eqnarray*}
\mathcal{W}_1(X_1,\mu_1,\nu_1;t)=\int\mathcal{W}(X_{1},\mu_{1},
\nu_{1};X_{2},\mu_{2},\nu_{2}; t)dX_{2}, \\
\mathcal{W}_2(X_2,\mu_2,\nu_2;t)=\int\mathcal{W}(X_{1},\mu_{1},\nu_{1};
X_{2},\mu_{2},\nu_{2}; t)dX_{1}\ .
\end{eqnarray*}
This condition can be used to distinguish an entangled state and will be discussed later.

The optical tomogram is related to the symplectic tomogram:
\[
\mathcal{W}_{0}(X_{1},\theta_{1},X_{2},\theta_{2})=\mathcal{W}(X_{1},
\cos\theta_{1},\sin\theta_{1},X_{2},\cos\theta_{2},\sin\theta_{2})\ ,
\]
which measures the quadratures in a rotated reference frame. Thus,
all the information of a quantum state is contained in the optical
(or symplectic) tomogram. The importance of the optical tomogram
consists in the fact that it can be obtained through homodyne
measurements for various systems~\cite{vogel}.

If the two-mode state is a Gaussian one, the symplectic tomogram
is described by a normal probability distribution:
\begin{eqnarray}
\mathcal{W}(X_{1},\mu_{1},\nu_{1};X_{2},\mu_{2},\nu_{2}; t)=
\frac{1}{2 \pi \sqrt{\det \boldsymbol{\sigma}_{XX}(t)}}
\exp\left(-\frac{1}{2}(X_{1}^{\prime},\,X_{2}^{\prime})\boldsymbol{\sigma}_{XX}^{-1}(t)
\left(\begin{array}{c}
X_{1}^{\prime}\\
X_{2}^{\prime}
\end{array}\right)\right),
\label{30}
\end{eqnarray}
where
\[
X_{1}^{\prime}=X_{1}-\left\langle X_{1}\right\rangle ,\quad X_{2}^{\prime}=X_{2}-\left\langle X_{2}\right\rangle\ ,
\]
and the dispersion matrix $\boldsymbol{\sigma}_{XX}(t)$ reads
\begin{equation}
\boldsymbol{\sigma}_{XX}(t)=\left(\begin{array}{cc}\sigma_{X_{1}X_{1}}(t) & \sigma_{X_{1}X_{2}}(t)\\
\sigma_{X_{2}X_{1}}(t) & \sigma_{X_{2}X_{2}}(t)
\end{array}\right).
\label{31}
\end{equation}
The mean values $\left\langle X_{1}\right\rangle $ and $\left\langle X_{2}\right\rangle $
are expressed in terms of the corresponding expectation values of the quadrature components of both modes:
\[
\begin{array}{c}
\left\langle X_{1}\right\rangle =\mu_{1}\left\langle \hat{q}_{1}\right\rangle +\nu_{1}\left\langle \hat{p}_{1}\right\rangle ,\quad
\left\langle X_{2}\right\rangle =\mu_{2}\left\langle \hat{q}_{2}\right\rangle +\nu_{2}\left\langle \hat{p}_{2}\right\rangle .
\end{array}
\]
The dispersions and covariance are
   \begin{eqnarray}
      \sigma_{X_{1}X_{1}}(t)&=&\mu_{1}^{2}\sigma_{q_{1}q_{1}}(t)+\nu_{1}^{2}\sigma_{p_{1}p_{1}}(t)+2\mu_{1}\nu_{1}\sigma_{q_{1}p_{1}}(t)\ ,\nonumber \\
      \sigma_{X_{2}X_{2}}(t)&=&\mu_{2}^{2}\sigma_{q_{2}q_{2}}(t)+\nu_{2}^{2}\sigma_{p_{2}p_{2}}(t)+2\mu_{2}\nu_{2}\sigma_{q_{2}p_{2}}(t)\ ,\nonumber \\
      \sigma_{X_{1}X_{2}}(t)&=&\mu_{1}\mu_{2}\sigma_{q_{1}q_{2}}(t)+\nu_{1}\nu_{2}\sigma_{p_{1}p_{2}}(t)+\mu_{1}\nu_{2}\sigma_{q_{1}p_{2}}(t)+\mu_{2}\nu_{1}\sigma_{q_{2}p_{1}}(t)\ .
\label{32}
   \end{eqnarray}
The Gaussian state is completely determined by the
covariance matrix and the mean values of the homodyne quadratures
also in the tomographic representation. The time evolution in the
Hamiltonian~(\ref{1}) of a Gaussian state is also a Gaussian state,
as it can be seen in Eq.~(\ref{11}), so the time evolution of the
tomogram in this system is given by Eq.~(\ref{30}).

The time-dependent functions: $\left\langle \hat{q}_{1}\right\rangle
$, $\left\langle \hat{q}_{2}\right\rangle $, $\left\langle
\hat{p}_{1}\right\rangle $, $\left\langle \hat{p}_{2}\right\rangle
$, $\sigma_{q_{1}q_{1}}$, $\sigma_{p_{1}p_{2}}$,
$\sigma_{q_{1}p_{1}}$, $\sigma_{q_{2}q_{2}}$, $\sigma_{p_{2}p_{2}}$,
$\sigma_{q_{2}p_{2}}$ are calculated in terms of the corresponding
wave function in the standard form. Therefore, to calculate the optical
and symplectic tomograms, we have to use the corresponding matrices
$\mathbf{A}_G$ and $\mathbf{B}_G$ for the different initial states
considered in this work, of course, by substituting properly the
matrices $\boldsymbol{\lambda}_k$ with $k=1,2,3,4$ carrying the
information of the evolution under the parametric amplifier.

\begin{figure}
\centering
\includegraphics[scale=0.3]{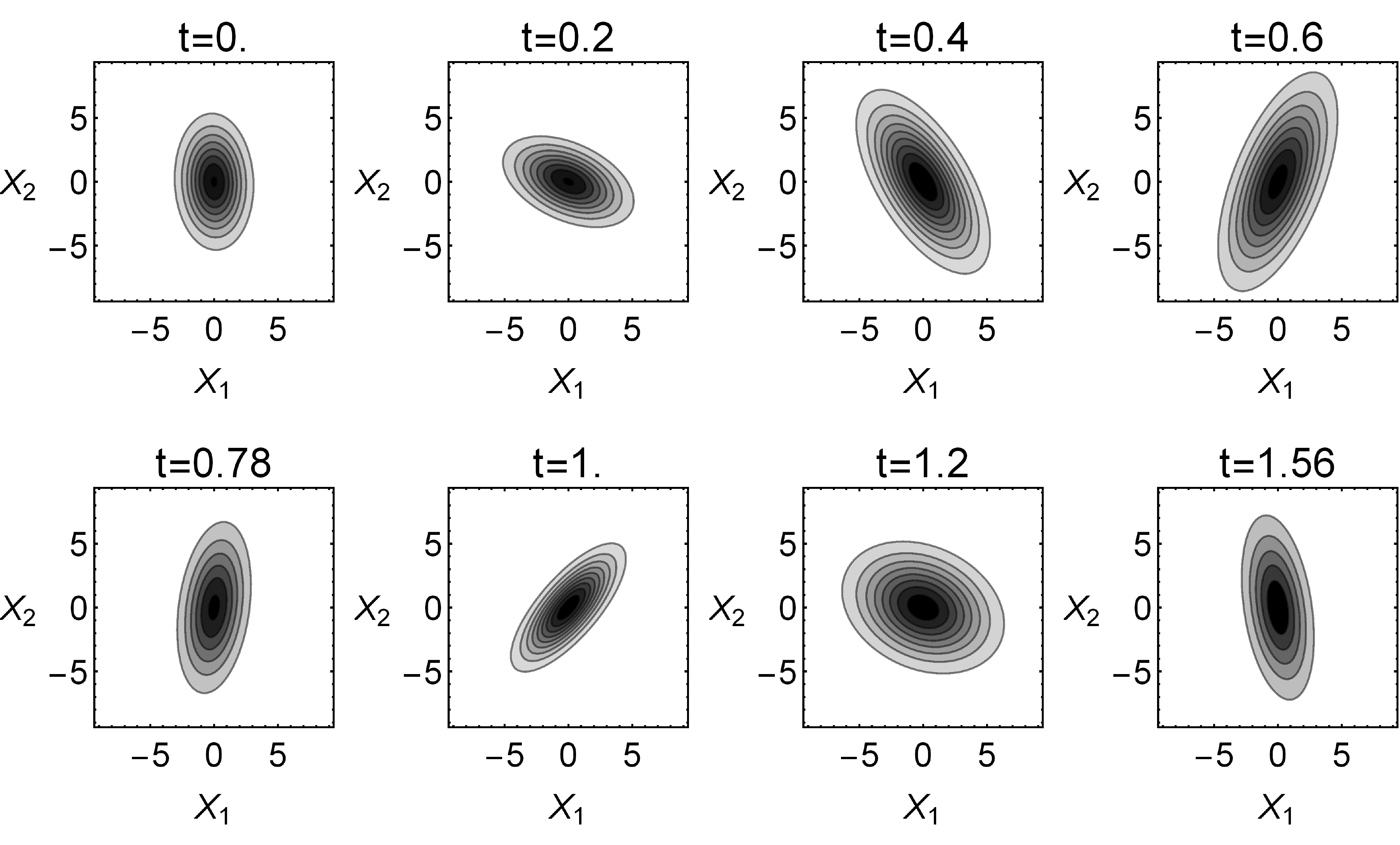}
%\captionsetup{justification=centering}
\caption{Contour plots for the tomogram in the phase space for the squeezed
vacuum state with $\beta=4/5$ at different times. We used for
the parametric amplifier $\Omega=9$, $k=2$, $\omega_a=1$, and
$\omega_b=3$. The tomogram parameters are $\mu_1=\cos \pi / 4, \nu_1=\sin \pi
/ 4$, $\mu_2=\cos \pi / 8, \nu_2=\sin \pi /8$.}
\label{fig:comprimido}
\end{figure}

In Fig.~\ref{fig:comprimido}, the evolution of the tomogram for the squeezed vacuum state in the parametric amplifier is presented.
The figure shows that at times $\pi / \nu$ and $2 \pi / \nu$ the
tomogram is not the same as the one at $t=0$; in fact, the
covariance matrix at those times is different. Although the
entanglement properties are the same at those times. Making use of the squeeze parameter $\eta
(t)$, one can check that $ \eta (n \pi / \nu) =e^{-i n \pi \omega
/\nu }\beta=- e^{-i n \pi \omega /\nu } \textrm{tanh} r$,  implying
that the state and its covariance matrix are
different.

\begin{figure}
\centering
\includegraphics[scale=0.3]{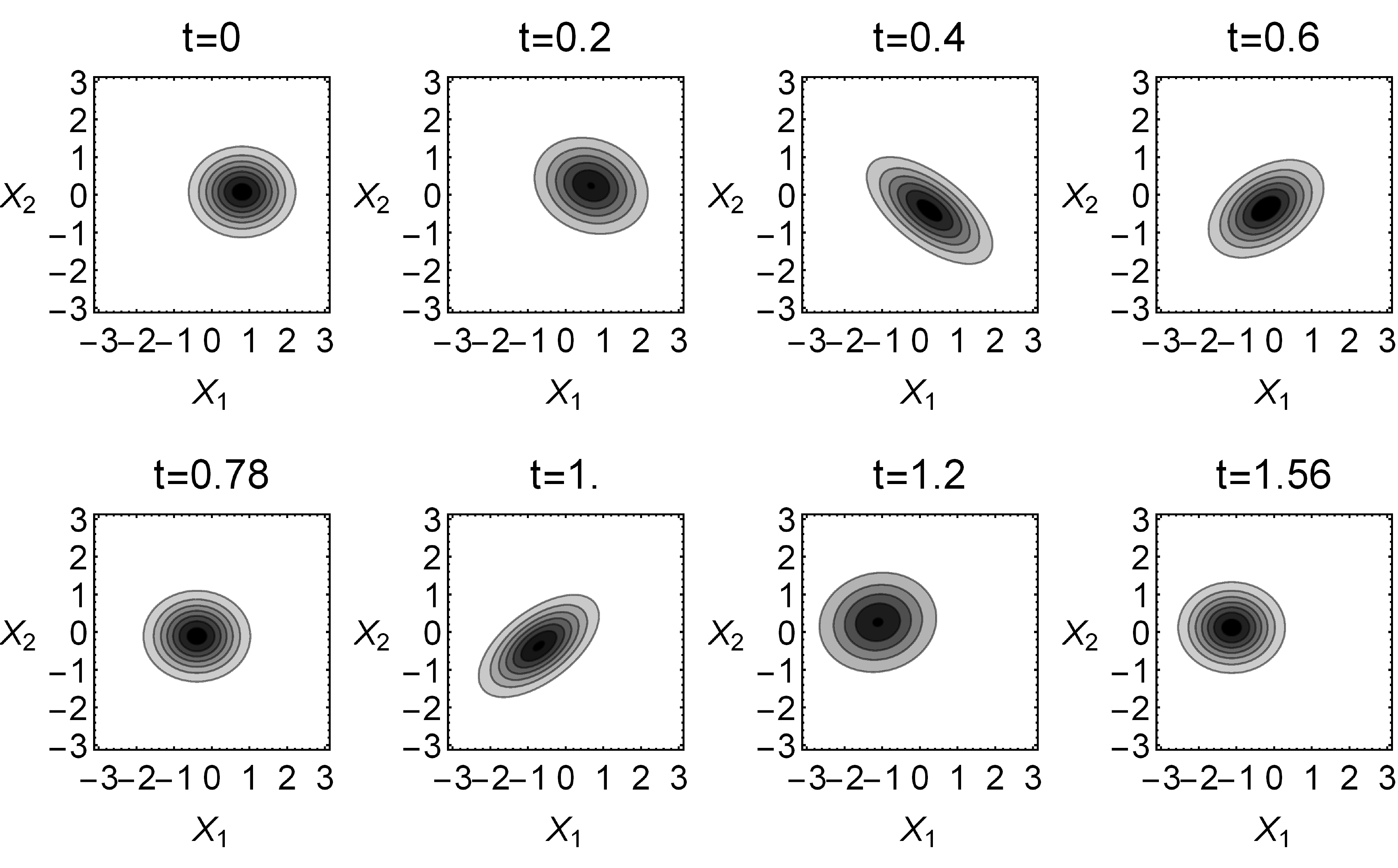}
\centering
\caption{Contour plots for the tomogram in the phase space for the coherent
state with $\alpha_1=4/5$, $\alpha_2=1/10$ at different times. The
parameters of the parametric amplifier are the same as in
Fig.~\ref{fig:comprimido} with the tomogram parameters are $\mu_1=\cos
\pi / 4, \nu_1=\sin \pi / 4$ and $\mu_2=\cos \pi / 8, \nu_2=\sin \pi
/8$.}
\label{fig:coherente}
\end{figure}

In Fig.~\ref{fig:coherente}, the tomograms for the coherent state
are displayed. The center of the wave packet moves according to the
mean values of the position operators but the shape of the tomogram
is the same at times $\pi / \nu$ and $2 \pi / \nu$. Additionally,
one can calculate the correlation between the two variables $X_1$
and $X_2$ represented by $\sigma_{X_1 X_2}$, and it is zero at those
times.

% Figura 1

% Figura 2

\section{Discretization of the density matrix, von Neumann and linear entropies}

The entanglement between two input modes in the parametric amplifier has been shown to exist through different methods, as second order correlations \cite{drummond,rekdal} and the calculation of the Duan {\it et al.} criterion \cite{fang}. The entanglement between two modes in a symmetric Gaussian state have been described through EPR inequalities and the von Neumann entropy in \cite{giedke}. The entanglement present in the parametric amplifier has been used in quantum metrology \cite{hudelist} and entanglement swapping \cite{zhang}. Also the entanglement in other parametric processes, related to the amplification as the parametric oscillation is frequently used to generate correlated light \cite{reid, cassemiro1, cassemiro2} and implement quantum information protocols \cite{jing,takei}.
The von Neumann and linear entropies are generally used to measure
the entanglement between the modes of a bipartite pure system. However
the analytic calculation of these quantities is, in general, not an
easy matter. In this section, we provide a numerical method to
calculate both entropies for a continuous variable density matrix
using a discrete form. This method also can be used to define a
positive map between the density matrix and other sub-matrices
similar to the reduced density matrices that retain information on
the entanglement of the system.

In order to compare the results and to evaluate the entanglement
given by the tomographic representation discussed later, we
calculate the von Neumann and linear entropies for the two-variable
system.

The density matrix resulting from the time evolution of the Gaussian
state described by Eq. (\ref{11}) is a continuous function of the
coordinates
\begin{equation}
\rho (x^{\prime}_1, x^{\prime}_2, x_1, x_2,t)=\langle x^{\prime}_1,
x^{\prime}_2 \vert \rho (t) \vert x_1, x_2 \rangle= \psi^*
(x^{\prime}_1, x^{\prime}_2,t) \, \psi (x_1, x_2,t) \ .
\end{equation}

To determine the entanglement properties from a continuous variable
density matrix, the discrete form of the density operator is made.
Let us take four sets of discrete numbers along the axis that define
the density matrix variables, that is,
\begin{equation*}
 \{ x_{1_1}^{\prime}, x_{1_2}^{\prime}, \cdots, x_{1_{N}}^{\prime} \} \, , \:
\{x_{2_1}^{\prime}, x_{2_2}^{\prime}, \cdots, x_{2_{N}}^{\prime} \} \, , \
\{x_{1_1}, x_{1_2}, \cdots, x_{1_{N}}\}  \, , \
\{x_{2_1}, x_{2_2}, \cdots, x_{2_{N}}\}  \, ,
\end{equation*}
where the size of the steps is $\Delta x_1=x_{1_{r+1}}-x_{1_{r}}$
and $\Delta x_2=x_{2_{r+1}}-x_{2_{r}}$. One can notice that, to
define properly the transpose matrix, one should take the same
number of elements for the coordinates $x_i$ and $x'_i$ and, for
simplicity, let us choose the same step between them: $\Delta
x_1^{\prime}=\Delta x_1$ and $\Delta x_2^{\prime}=\Delta x_2$. These
partitions must be chosen to guarantee the normalization condition
of the density matrix.

Therefore, the discrete two-mode density matrix can be expressed as
\begin{equation*}
\rho_{i,j,k,l}(t)=\rho (x_{1_i}^{\prime},x_{2_j}^{\prime},x_{1_k},x_{2_l},t),
\end{equation*}
where the normalization condition is expressed in the form
\begin{equation*}
\sum_{i,j=1}^{N} \rho_{i,j,i,j} (t) \Delta x_1 \Delta x_2 =1 \ .
\end{equation*}
Then, the corresponding definition of the partial density matrix of
the mode~1 is given by
\begin{equation*}
\rho_{i,j}^{(1)}(t)=\sum_{k=1}^{N} \rho_{i,k,j,k} (t) \Delta x_2 \ ,
\end{equation*}
obtaining the eigenvalues of the reduced density matrix provide us with a method to calculate either the linear or the von Neumann entropies.

The eigenvectors and eigenvalues are obtained by solving the
standard eigenvalue equation for the matrix $\rho^{(1)}_{ij}(t)
\Delta x_1$. Denoting the corresponding eigenvalues as $\{e_{k}
(t)\}$, the linear and von Neumann entropies can be calculated as
\begin{equation}
S_L=1-\sum_{k=1}^{N} e_{k}^2 (t),\qquad S_{VN}=-\sum_{k=1}^{N} e_{k}
(t) \ln (e_k(t)) \ .
\end{equation}

For the squeezed vacuum state the linear entropy is given by
\begin{equation}
S_{L}(t)=\frac{2 \left|\eta (t)\right|^{2}}{1+\left|\eta (t) \right|^{2}}\ ,
\label{27}
\end{equation}
with $\eta$ given by Eq.~(\ref{21}). Similarly, for the von Neumann
entropy, one gets
\begin{equation}
S_{VN}(t)=-\ln\left(1-\left|\eta (t)\right|^{2}\right)-\frac{\left|
\eta (t)\right|^{2}\ln\left(\left|\eta (t)\right|^{2}\right)}{1-\left|
\eta (t)\right|^{2}}\ .
\label{28}
\end{equation}

%Figura 3
\begin{figure}
\centering
\includegraphics[scale=0.3]{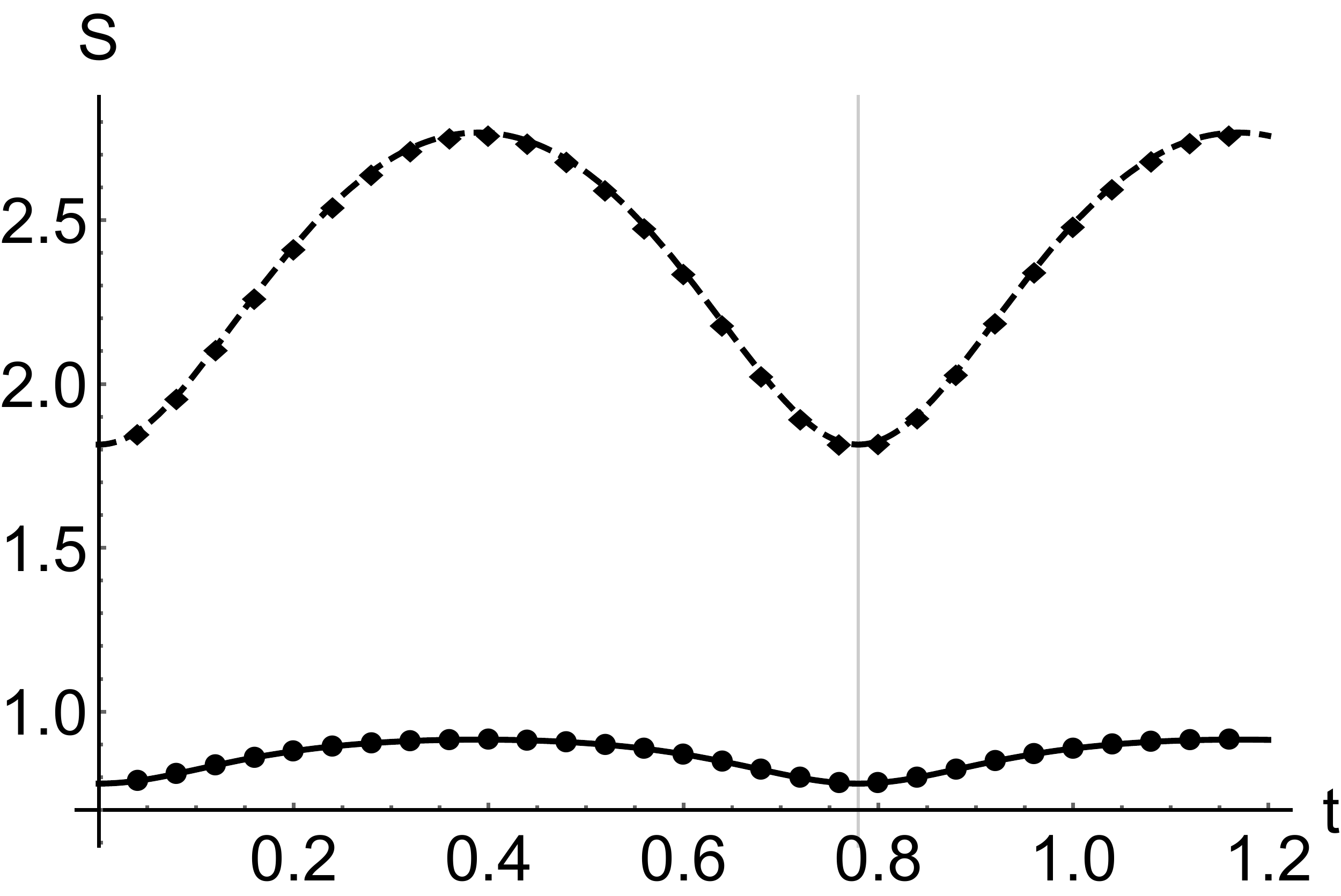}
\includegraphics[scale=0.3]{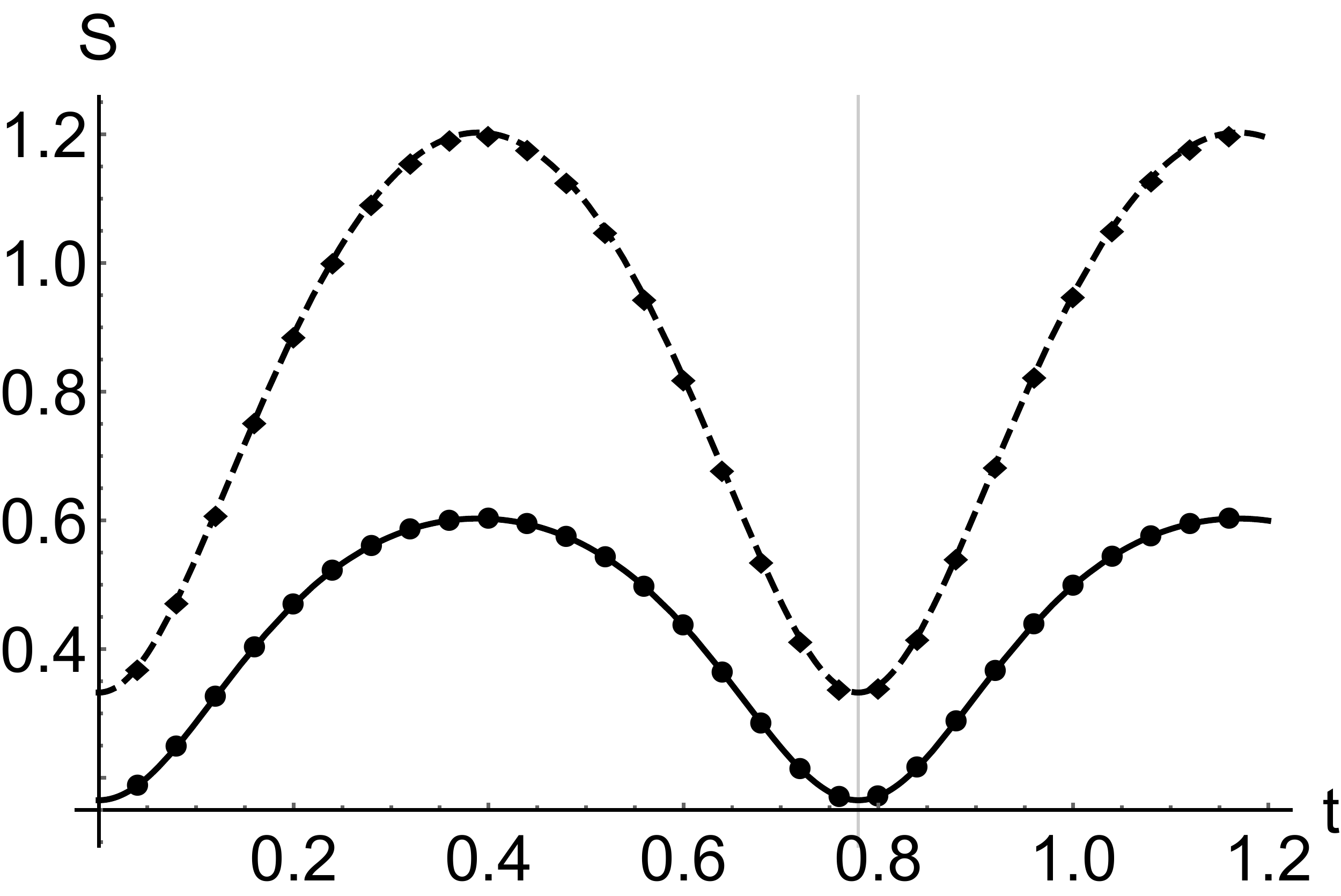}
\caption{The analytic results for the linear entropy (solid black) and
von Neumann entropy (dashed black) as functions of time for the
squeezed vacuum state with the squeeze parameter (left) $\beta=4/5$
and (right) $\beta=3/10$. For the parametric amplifier, we take
$\Omega=9$, $k=2$, $\omega_a=1$, and $\omega_b=3$. The corresponding
numerical results are also shown in this figure where the linear
entropy is indicated by a black dotted curve while the von Neumann
entropy is displayed by a black rhombus curve. }
\label{fig:entropies1}
\end{figure}
In Fig.~\ref{fig:entropies1}, we compare the analytic and numerical
results for the von Neumann and linear entropies for a squeezed
vacuum state evolving in the parametric amplifier. The difference
between the analytic and numerical results has a maximum value of
$10^{-6}$. In this figure, one can see that the same entanglement is
obtained for times $t=0$ and $t=\pi / \nu$, as both entropies show a
periodic behavior with period $T=\pi / \nu$.

% Figura 4
\begin{figure}
\centering
\includegraphics[scale=0.30]{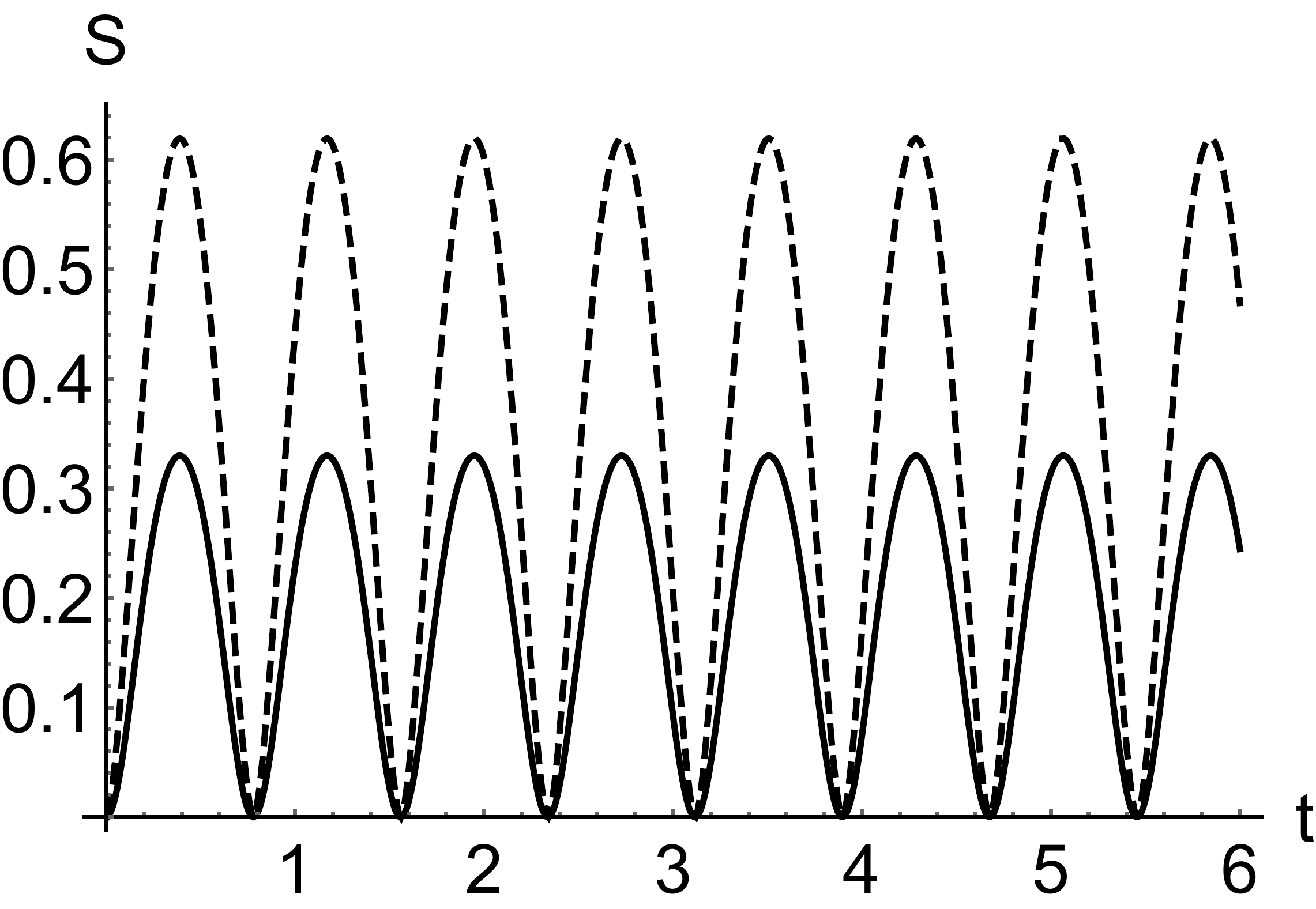}
\includegraphics[scale=0.30]{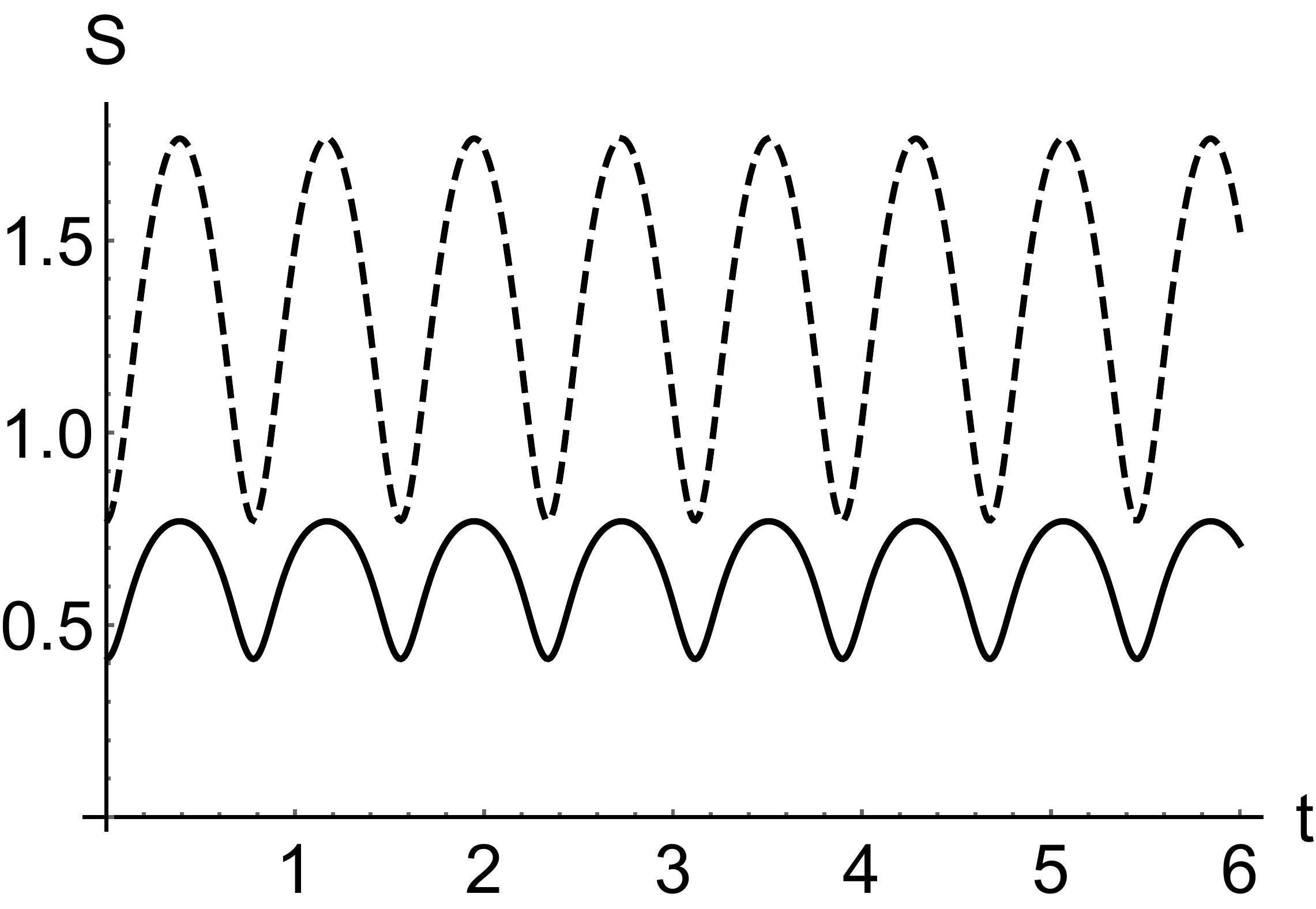}
\caption {The linear entropy (solid curve) and von Neumann entropy (dashed curve)
as functions of time for the coherent state with parameters
$\alpha_1=1$, $\alpha_2=3$~(left) and for the particular Gaussian
state with $a_{11}=1$, $a_{22}=3$, and $ a_{12}=1.4$~(right). We use
the same parameter values for the parametric amplifier as in
Fig.~\ref{fig:entropies1}.}
\label{fig:entropies2}
\end{figure}

Also in Fig~\ref{fig:entropies2}, the von Neumann and linear
entropies for the coherent state and a particular Gaussian state are
presented. We have used the values $a_{11}=1$, $a_{22}=3$, $
a_{12}=1.4$, $\Omega=9$, $k=2$, $\nu=\sqrt{\Omega^{2}/4
-k^{2}}=\sqrt{65}/2$, $\omega_a=1$, and $\omega_b=3$. We can observe
again an oscillatory behavior of period $T=\pi/\nu$.  Also one can
see that in the coherent state the two modes initially are not
entangled but the evolution in the parametric amplifier makes these
two modes entangled.

The oscillation of the entropies imply that there is entanglement between the two Gaussian modes even when initially are not entangled due to the evolution in the parametric amplifier. This entanglement has a maximum value at times $t=n \pi /(2 \nu)$ for $n$ odd and has a minimum value at times $t=n \pi / \nu$ with $n$ even. We note that the minimum value for the entanglement is equal to the initial entropy, so given an initial entangled state the evolution in the amplifier can increase the entanglement between the modes.

\section{Qubit portrait of symplectic tomograms}

In this section, the qubit portrait of a symplectic tomogram is
defined and calculated, in particular, for two-mode pure Gaussian
states evolving in the parametric amplifier. This qubit portrait is
the reduction of a symplectic or optical tomograms to a $4$
component probability vector, which for simply separable states is
the tensor product of two-dimensional probability vectors.

This qubit portrait can be used to determine a Bell-type inequality
where the violation of the parameter $\vert\mathcal{B}\vert\leq 2$
indicates that the bipartite system cannot be separable, i.e., it is
entangled. Now, the converse statement is not true.

The spin tomograms for time-dependent Hamiltonians linear in spin
variables have been constructed in~\cite{castanos2}.  These
tomograms have been also used to study qubits and qudits within the
quantum information context of separable and entangled
states~\cite{sudarshan}.  To define the qubit portrait for a
continuous variable system, we generalize the idea developed for the
spin tomogram, for which a qubit portrait of qudit states and
Bell-type inequalities have been proposed in~\cite{chernega}.

The spin states can be described by a probability distribution
called spin tomogram denoted by $\omega(m, \vec{n})$, where $m$ is
the projection in the direction $\vec{n}$~\cite{dodonov}. In
general, the tomogram of a $d$-dimensional qudit system has $2^d$
components corresponding to the different projections of the angular
momentum operator. The qubit portrait is defined as the reduction of
the tomogram of a qudit system to a two-qubit tomogram. To construct
the portrait, we reduce the $2^d$ components to only 4; to make
this, we construct 4 arbitrary sets of this components and sum them.

In an analogous form, one can reduce all the information contained
in the symplectic or optical tomograms to $4$ numbers related with
the probabilities to find the quadrature components $X_1$ and $X_2$
into the $4$ integration regions
($\mathbb{A}_1,\mathbb{A}_2,\mathbb{A}_3,\mathbb{A}_4$), that do not
overlap $\mathbb{A}_i\cap \mathbb{A}_j=\emptyset$; and the union of
all these regions is equal to the complete two-dimensional space
$\mathbb{R}^2$. Then each components of the four-dimensional
probability vector will be given by
\begin{equation}
P_{i}(\boldsymbol{\mu}_1, \boldsymbol{\mu}_2)=\int_{\mathbb{A}_{i}}\mathcal{W}(X_{1},X_2,\mu_{1},
\nu_{1},\mu_{2},\nu_{2}; t)dX_{1}dX_2,
\label{prob}
\end{equation}
with $i=1,\cdots,4$ and $\boldsymbol{\mu}_k=(\mu_k,\nu_k)$.

Then, we define a $4\times 4$ stochastic matrix $\mathbf{M}$ as
follows:
\begin{equation}
\mathbf{M}=\left(\begin{array}{cccc}
P_{1}(\boldsymbol{\mu}_a,\boldsymbol{\mu}_b) & P_{1}(\boldsymbol{\mu}_a,\boldsymbol{\mu}_c) & P_{1}(\boldsymbol{\mu}_d,\boldsymbol{\mu}_b) & P_{1}(\boldsymbol{\mu}_d,\boldsymbol{\mu}_c)\\
P_{2}(\boldsymbol{\mu}_a,\boldsymbol{\mu}_b) & P_{2}(\boldsymbol{\mu}_a,\boldsymbol{\mu}_c) & P_{2}(\boldsymbol{\mu}_d,\boldsymbol{\mu}_b) & P_{2}(\boldsymbol{\mu}_d,\boldsymbol{\mu}_c)\\
P_{3}(\boldsymbol{\mu}_a,\boldsymbol{\mu}_b) & P_{3}(\boldsymbol{\mu}_a,\boldsymbol{\mu}_c) & P_{3}(\boldsymbol{\mu}_d,\boldsymbol{\mu}_b) & P_{3}(\boldsymbol{\mu}_d,\boldsymbol{\mu}_c)\\
P_{4}(\boldsymbol{\mu}_a,\boldsymbol{\mu}_b) & P_{4}(\boldsymbol{\mu}_a,\boldsymbol{\mu}_c) & P_{4}(\boldsymbol{\mu}_d,\boldsymbol{\mu}_b) & P_{4}(\boldsymbol{\mu}_d,\boldsymbol{\mu}_c)
\end{array}\right).
\label{39}
\end{equation}
Here, each column vector specify the two-dimensional coordinate
system where the measurements of the position operators are
realized. Each one of them satisfy $\sum_k P_{k}(\boldsymbol{\mu}_l, \boldsymbol{\mu}_k)=1$.
It can be shown that in the case of a simply separable state, the
matrix $\mathbf{M}$ can be written as the direct product of two
subsystems. In this case, one can define a Bell-type inequality.

Let us consider two stochastic matrices $\left(\begin{array}{cc}
x & y\\
1-x & 1-y
\end{array}\right)$,$\quad\left(\begin{array}{cc}
t & z\\
1-t & 1-z
\end{array}\right)$, and their tensor product
\begin{equation}
\widetilde{\mathbf{M}}=\left(\begin{array}{cc}
x & y\\
1-x & 1-y
\end{array}\right)\otimes\left(\begin{array}{cc}
t & z\\
1-t & 1-z
\end{array}\right)\, .
\label{mtilde}
\end{equation}
It is straightforward to show that the matrix elements of $\widetilde{M}_{jk}(x,y,z,t)$
satisfy the Laplace equation
\[
\left(\frac{\partial^2}{\partial x^{2}}+\frac{\partial^2}{\partial y^{2}}
+\frac{\partial^2}{\partial z^{2}}+\frac{\partial^2}{\partial t^{2}}\right)
\widetilde{M}_{jk}(x,y,z,t)=0\, .
\]
This means that the extreme values of any function
\begin{equation}
\mathcal{B}(x,y,z,t)=\sum_{j,k=1}^{4} \widetilde{M}_{jk}(x,y,z,t) \, {C}_{kj} \label{36}
\end{equation}
are situated on the boundaries of the region where $x,y,z,t$ are
given. In our case, it is the cube $0\leq x,y,z,t\leq1$. For this work we have taken the
coefficient $C_{kj}$ in the matrix form
\[
\mathbf{C}=\left(\begin{array}{cccc}
1 & -1 & -1 & 1\\
1 & -1 & -1 & 1\\
1 & -1 & -1 & 1\\
-1 & 1 & 1 & -1
\end{array}\right),
\]
although the election of $\mathbf{C}$ is not unique, this matrix had been chosen in order to obtain a function $\mathcal{B}$ with two elements from each column of $\widetilde{\mathbf{M}}$ with a positive sign and two elements with negative sign. In the case where all this elements cancel each other the parameter $\mathcal{B}$ would be equal to zero and in general one can check that the following inequality holds:
\begin{equation}
|\mathcal{B}(x,y,z,t)|\leq2 \ . \label{37}
\end{equation}

Using the stochastic matrix $\mathbf{M}$, as in Eq. (\ref{mtilde}), and taking
the definition of the Bell-type parameter, as in Eq. (\ref{36}), one
can evaluate the inequality
\begin{equation}
\vert \mathcal{B}\vert=\vert E(a,b)+E(a,c)+E(d,b)-E(d,c) \vert \leq 2
\end{equation}
where
$E(x,y)=P_1(\boldsymbol{\mu}_x,\boldsymbol{\mu}_y)-P_2(\boldsymbol{\mu}_x,\boldsymbol{\mu}_y)-P_3(\boldsymbol{\mu}_x,\boldsymbol{\mu}_y)
+P_4(\boldsymbol{\mu}_x,\boldsymbol{\mu}_y)$. This inequality evaluates, if the matrix
$\mathbf{M}$ can be expressed as a direct product of two subsystems,
all the separable states must satisfy this condition. Therefore, a
violation of this inequality is a sufficient condition for
entanglement.

To establish properly the Bell-type inequality, the integrating
regions used in the construction of the matrix $\mathbf{M}$ should
be taken as the direct product of the two regions in $X_1$ and $X_2$
in order to preserve the product structure of the matrix
$\mathbf{M}$, this is,
\begin{equation}
\left( \begin{array}{c}
\mathbb{A}_1 \\
\mathbb{A}_2 \\
\mathbb{A}_3 \\
\mathbb{A}_4
\end{array}\right)=\left(
\begin{array}{c}
\mathcal{L}_1^{(1)} \\
\mathcal{L}_2^{(1)} \end{array}\right) \otimes
\left(
\begin{array}{c}
\mathcal{L}_1^{(2)} \\
\mathcal{L}_2^{(2)} \end{array}\right) \ ,
\label{regions}
\end{equation}
where $\sum^4_{k=1}{\mathbb{A}_{k}}= \mathbb{R}^2$ and $\sum_{i=1}^2
\mathcal{L}_i^{(j)}=(-\infty,\infty)$, with $j=1,2$.
We enhance that the probabilities necessary to the definition of $\widetilde{\mathbf{M}}$ in Eq. (\ref{39}) can be obtained making use of a discrete scheme for the tomogram, similar to the one used for the density matrix in Section 5. In that scheme the probabilities are the discrete integral over the different areas $\mathbb{A}_1$ to $\mathbb{A}_4$. Using this discrete form and the fact that the optical tomogram can be observed experimentally \cite{bellini1} provide the possibility to measure this probabilities and to make measurements of the Bell inequalities previously studied.

\begin{figure}
\centering
\includegraphics[scale=0.265]{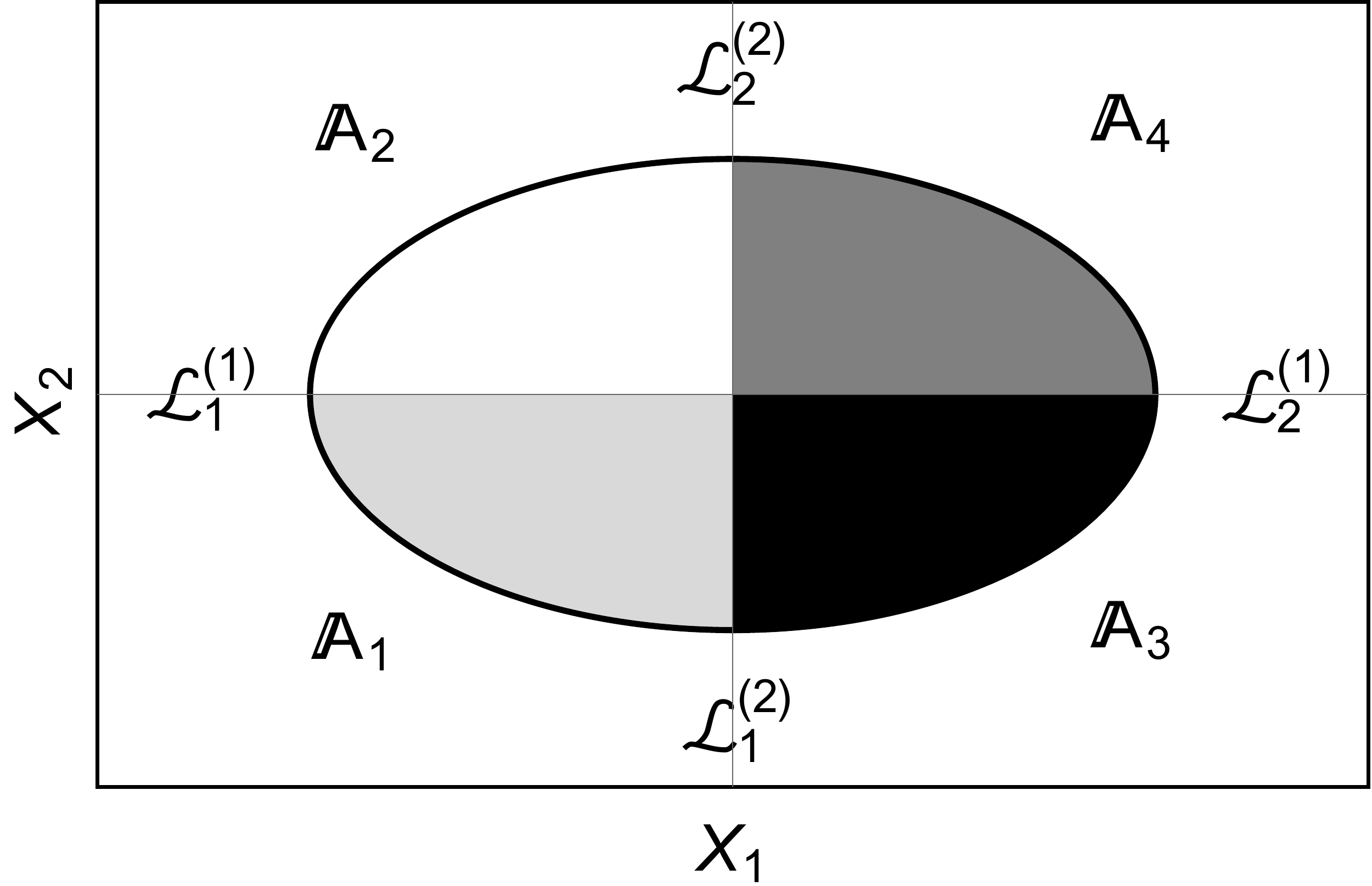} \hspace{5pt}
\includegraphics[scale=0.17]{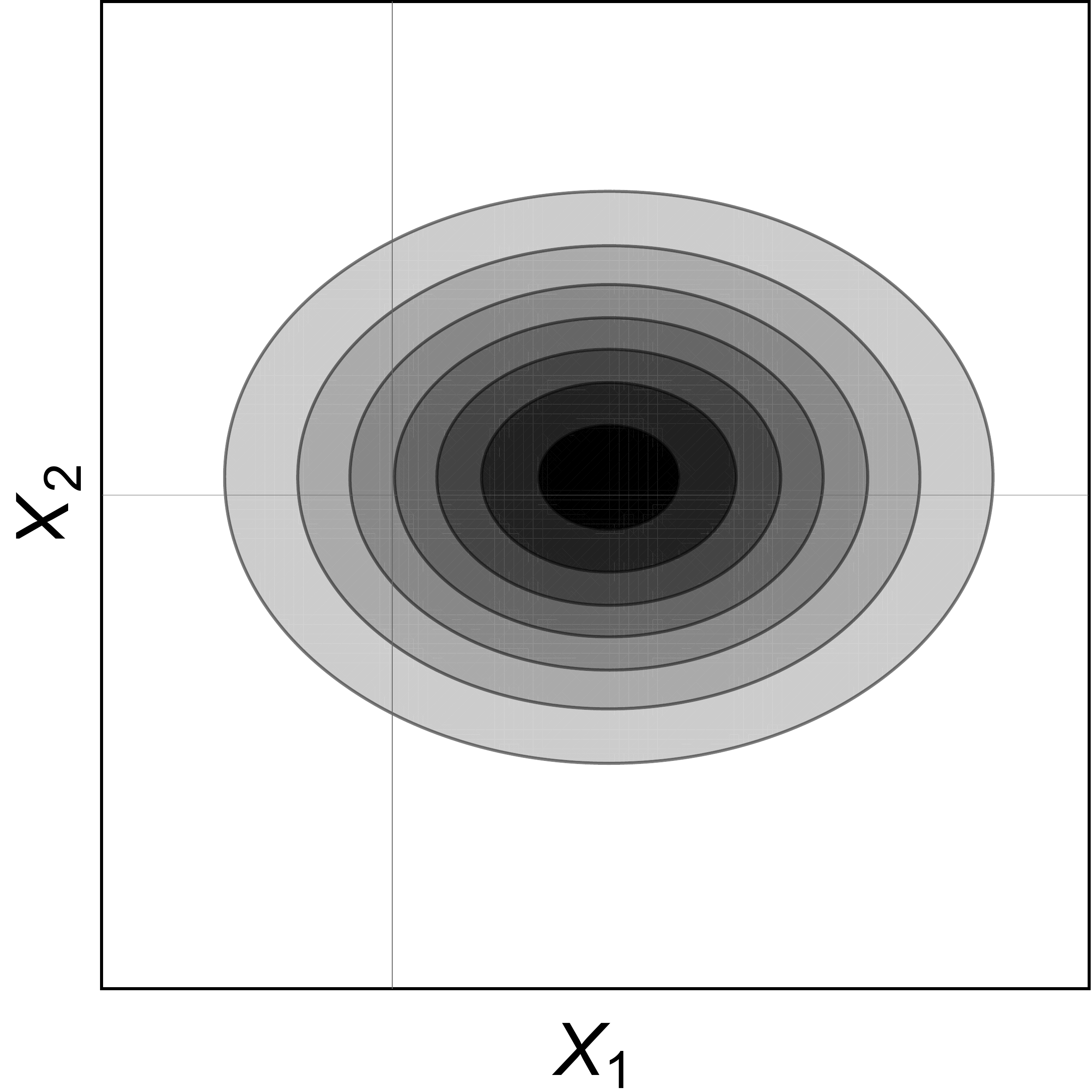}
\caption{(Left) Different regions $\mathbb{A}_1$ to $\mathbb{A}_4$ taken to define the probabilities in matrix $\mathbf{M}$, i.e. for a fixed $(\boldsymbol{\mu}_1, \boldsymbol{\mu}_2)$. The ellipse represents a contour plot of the tomogram and in the different colors are the areas corresponding to each of the four components in a column of $\mathbf{M}$. To compute the parameter $\mathcal{B}$ one need to obtain the corresponding areas for the other three values of $(\boldsymbol{\mu}_1, \boldsymbol{\mu}_2)$. (Right) Contour plot of the symplectic tomogram displaying the four different regions for a coherent state}
\label{fig:regions}
\end{figure}

In the present work, we consider
\begin{equation}
\mathcal{L}_1^{(j)}=(-\infty,0 ], \quad \mathcal{L}_2^{(j)}=[0, \infty),
\quad \textrm{with }j=1,2 \ ,
\end{equation}
these areas and line elements are represented in Fig.~\ref{fig:regions}.

% Figura 5

For the coordinate axes where the tomogram are measured, we used two different sets of
parameters indicated in Table~\ref{tabla}.

\begin{table}
\centering
\begin{tabular}{|c c||c c|}
\hline
\multicolumn{1}{|c}{$\boldsymbol{\mu}=(\mu,\nu)$} & $\theta$ &
\multicolumn{1}{c}{$\boldsymbol{\mu}=(\mu,\nu)$} & $\theta$\tabularnewline
\hline
${\boldsymbol{\mu}_{a}=(-0.39,\ -0.92})$ & $4.31\ (246.9^{\circ})$ &
${\boldsymbol{\mu}_{a}=(\quad 1,\  0})$ & 0\tabularnewline
%\hline
${\boldsymbol{\mu}_{b}=(-0.99,\ -0.01})$ & $3.15\ (180.5^{\circ})$ &
\ \ \quad ${\boldsymbol{\mu}_{b}=(0.92,\ 0.38})$ & $\pi/8$\tabularnewline
%\hline
${\boldsymbol{\mu}_{c}=(\ \ 0.02,\ \ \ 0.99})$ & $1.54\ (88.2^{\circ}) $ &
\ \ \quad ${\boldsymbol{\mu}_{c}=(0.38,\ 0.92})$ & $3 \pi/8$\tabularnewline
%\hline
${\boldsymbol{\mu}_{d}=(-0.60,\ -0.80})$ & $4.07\ (233.2^{\circ}) $ &
\ \ \ \qquad ${\boldsymbol{\mu}_{d}=(1/\sqrt{2},\ 1/\sqrt{2}})$ & $\pi/4$\tabularnewline
\hline
\end{tabular}
\caption{The parameters are given by $(\mu, \nu)=(s \cos\theta,s^{-1} \sin\theta)$
with $s=1$. The angles and scaling factors taking in the qubit portrait are indicated,
the selected values satisfy the constraint $2 \vert \mu \nu\vert \leq 1$.}
\label{tabla}
\end{table}

% Figura 6
\begin{figure}
\centering
\includegraphics[scale=0.30]{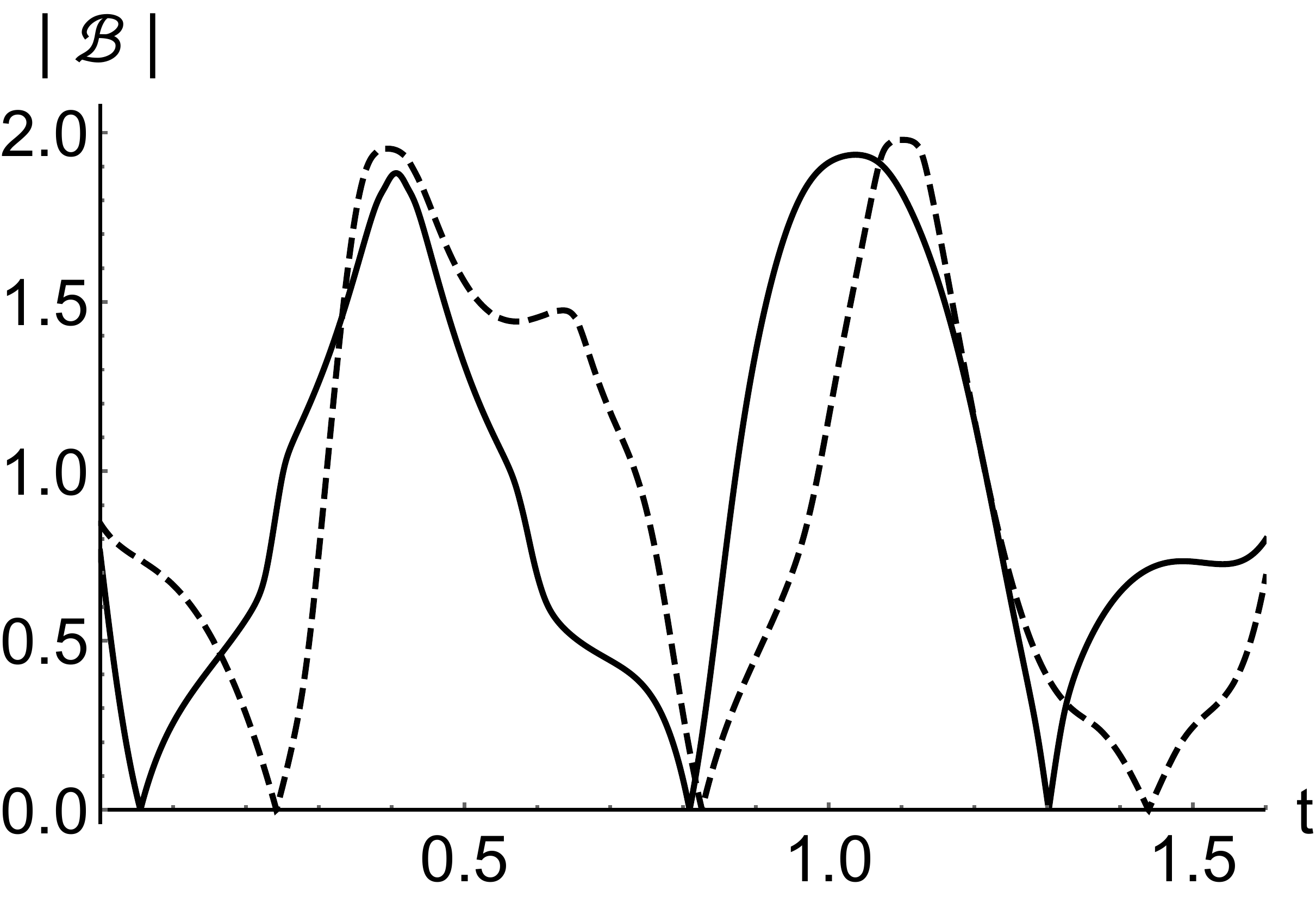} \quad
\includegraphics[scale=0.30]{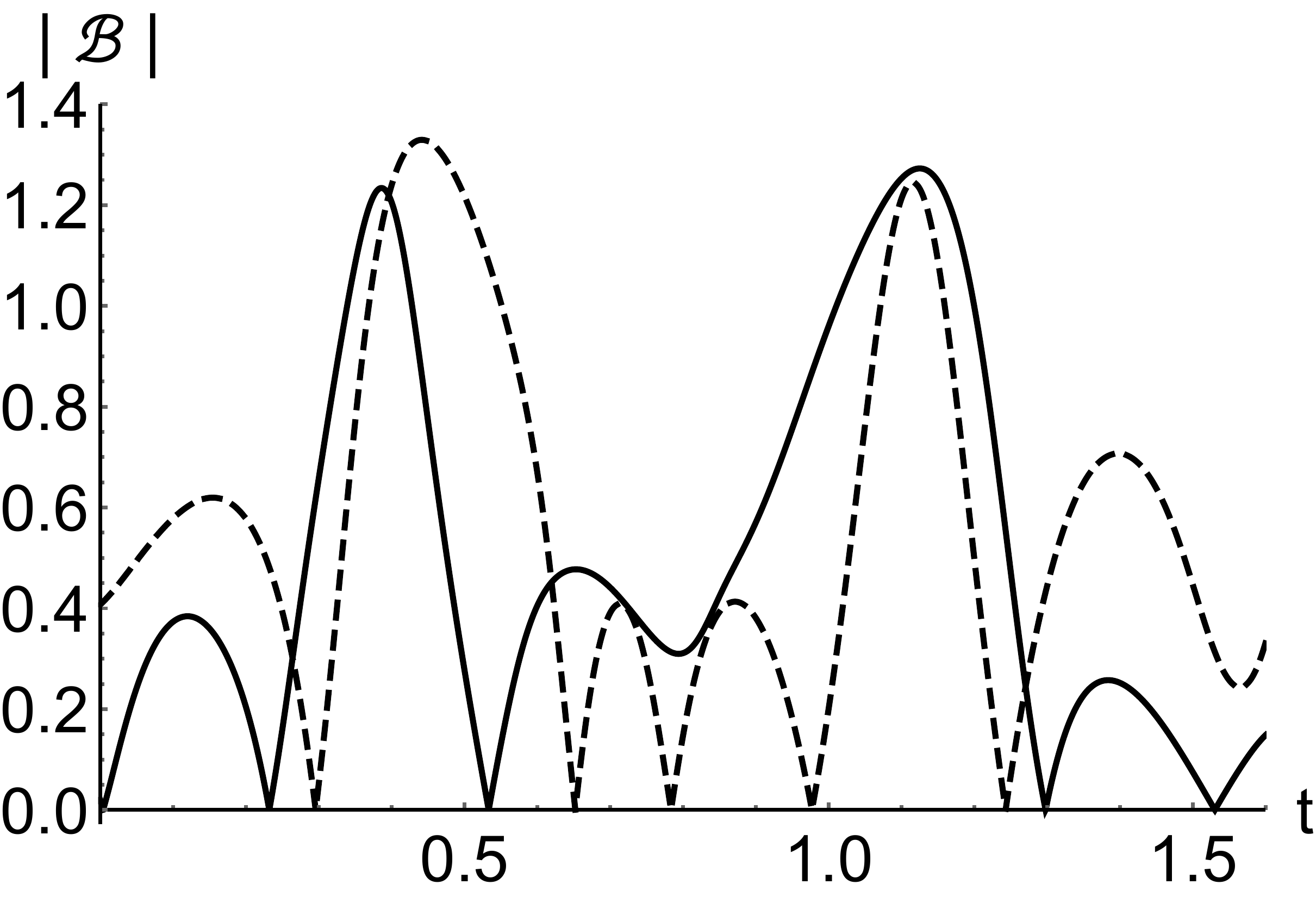}
\caption{Plot of the parameter $\vert \mathcal{B} \vert$ as a function of time
for the squeezed vacuum state with squeezing parameter
$\beta=4/5$~(left) and for the coherent state with parameters
$\alpha_1=4/5$ and $\alpha_2=1/10$~(right). Here, the amplifier
parameters are: $\omega_a=1$, $\omega_b=3$, $\Omega=9$, $k=2$, and
$\nu=\sqrt{65}/2$. In both cases, the solid black curve corresponds
to a system, where the parameters $\boldsymbol{\mu}$ are given at the left of
Table~ \ref{tabla}, and for the dashed curve, the parameters are
given at the right of Table~\ref{tabla}.}
\label{fig:vacio_bell}
\end{figure}

In Fig.~\ref{fig:vacio_bell}, the evaluation of the parameter
$\mathcal{B}$ is given for the squeezed vacuum state and the
coherent state, respectively. We did not found a violation of the
bound~2 for the parameter $\mathcal{B}$ in the two cases, although
the analysis was made for several partitions. It can be seen for the
non-continuous curve at times around $t= 0.3$ and $t=1$ that the
$\mathcal{B}$ parameter for the squeezed vacuum state is almost
zero. At these times, the contour plot of the tomogram of
Fig.~\ref{fig:comprimido} exhibits a small variance in one of the
principal axis of symmetry. In contrast to the times where the Bell
parameter $\mathcal{B}$ is near the value~2 (times around $0.5$ and
$1.2$), the tomogram displays a more symmetric distribution.

For the coherent state, the Bell parameter shows faster oscillations
than for the squeezed vacuum state and the local maxima are smaller.
This can be related to the displacement of the tomogram (see
Fig.~\ref{fig:coherente}), which implies also to the fact that the
variance, the mean values, and the probabilities defining the matrix
$\mathbf{M}$ are changing in a more complicated form.

\subsection{Alternative method}
Here, we want to propose an alternative form of using the portrait
tomogram of the Bell's parameter, which uses the fact that the
tomogram of a non-entangled two-dimensional state can be put as the
product of the reduced tomograms, as in Eq.~(\ref{separable}). The
difference of the probabilities $\int_{\mathbb{A}}
\mathcal{W}(1,2) dX_1 dX_2-\int_{\mathcal{L}^{(1)}} \mathcal{W}(1)
dX_1 \int_{\mathcal{L}^{(2)}} \mathcal{W}(2) dX_2$ is zero when the
system state is simply separable. This implies that the matrix
$\widetilde{\mathbf{M}}$ of the product of the reduced tomograms
$\mathcal{W}(1)\mathcal{W}(2)$ must be equal to the complete
two-variable tomogram $\mathcal{W}(1,2)$. Defining the matrix
$\widetilde{\mathbf{M}}$ of the product of the reduced tomograms as
in Eq.~(\ref{mtilde})
%\begin{equation}
%\fl
%\mathbf{\widetilde{M}}=\left(\begin{array}{cccc}
%\widetilde{P}_{1}(\boldsymbol{\mu}_a,\boldsymbol{\mu}_b) & \widetilde{P}_{1}(\boldsymbol{\mu}_a,\boldsymbol{\mu}_c) & \widetilde{P}_{1}(\boldsymbol{\mu}_d,\boldsymbol{\mu}_b) & \widetilde{P}_{1}(\boldsymbol{\mu}_d,\boldsymbol{\mu}_c)\\
%\widetilde{P}_{2}(\boldsymbol{\mu}_a,\boldsymbol{\mu}_b) & \widetilde{P}_{2}(\boldsymbol{\mu}_a,\boldsymbol{\mu}_c) & \widetilde{P}_{2}(\boldsymbol{\mu}_d,\boldsymbol{\mu}_b) & \widetilde{P}_{2}(\boldsymbol{\mu}_d,\boldsymbol{\mu}_c)\\
%\widetilde{P}_{3}(\boldsymbol{\mu}_a,\boldsymbol{\mu}_b) & \widetilde{P}_{3}(\boldsymbol{\mu}_a,\boldsymbol{\mu}_c) & \widetilde{P}_{3}(\boldsymbol{\mu}_d,\boldsymbol{\mu}_b) & \widetilde{P}_{3}(\boldsymbol{\mu}_d,\boldsymbol{\mu}_c)\\
%\widetilde{P}_{4}(\boldsymbol{\mu}_a,\boldsymbol{\mu}_b) & \widetilde{P}_{4}(\boldsymbol{\mu}_a,\boldsymbol{\mu}_c) & \widetilde{P}_{4}(\boldsymbol{\mu}_d,\boldsymbol{\mu}_b) & \widetilde{P}_{4}(\boldsymbol{\mu}_d,\boldsymbol{\mu}_c)
%\end{array}\right),
%\label{tomored}
%\end{equation}
where the probabilities are defined as
\begin{eqnarray}
x=\int_{\mathcal{L}^{(1)}_1} \mathcal{W}_1 (X_1,\mu_a,\nu_a) \ dX_1,
\quad y=\int_{\mathcal{L}^{(1)}_1} \mathcal{W}_1 (X_1,\mu_d,\nu_d) \ dX_1, \nonumber \\
t=\int_{\mathcal{L}^{(2)}_1} \mathcal{W}_2 (X_2,\mu_b,\nu_b) \ dX_2,
\quad z=\int_{\mathcal{L}^{(2)}_1} \mathcal{W}_2 (X_2,\mu_c,\nu_c) \
dX_2,
\end{eqnarray}
%\begin{eqnarray}
%\widetilde{E}_{i k}(\boldsymbol{\mu}_1, \boldsymbol{\mu}_2)=\int_{\mathcal{L}^{(1)}_i}\mathcal{W}_1(X_{1},\mu_{1},\nu_{1}; t)dX_1 \int_{\mathcal{L}^{(2)}_k}\mathcal{W}_2(X_2,\mu_{2},\nu_{2}; t)dX_2, \nonumber \\
%\end{eqnarray}
%as $\widetilde{P}_1=\widetilde{E}_{11}$, $\widetilde{P}_2=\widetilde{E}_{12}$, $\widetilde{P}_3=\widetilde{E}_{21}$ and $\widetilde{P}_4=\widetilde{E}_{22}$.
then a parameter $\widetilde{\mathcal{B}}$ corresponding to this
matrix $\widetilde{\mathbf{M}}$ can be defined in the form
\begin{equation}
\widetilde{\mathcal{B}}=\sum_{j,k=1}^4 \widetilde{M}_{jk} C_{kj}\ .
\end{equation}

For non-entangled pure states, the subtraction of the parameter for
the complete tomogram $\mathcal{B}$ and the product of the reduced
tomograms $\widetilde{\mathcal{B}}$ must satisfy the equality
\begin{equation}
\vert \mathcal{B}-\widetilde{\mathcal{B}} \vert=0\ ,
\label{nonentangled}
\end{equation}
while the entangled states satisfy the inequality
\begin{equation}
\vert \mathcal{B}-\widetilde{\mathcal{B}} \vert>0\ .
\label{entangled}
\end{equation}

The condition (\ref{entangled}) can be used to distinguish entangled
from non-entangled states but the converse statement is not always
fulfilled, i.e., the condition $\vert
\mathcal{B}-\widetilde{\mathcal{B}} \vert=0$ does not imply that the
system state is separable. Therefore, the expression in Eq.
(\ref{nonentangled}) is a necessary condition for the system state
to be separable and a violation to this equality is a sufficient
condition for entanglement.

The quantity $\widetilde{\mathcal{B}}$ can be measured experimentally in the discrete scheme of the tomogram, making possible the evaluation of  the separability criteria $\vert \mathcal{B}-\widetilde{\mathcal{B}} \vert$ for these systems.

% Figura 7
\begin{figure}
\centering
\includegraphics[scale=0.24]{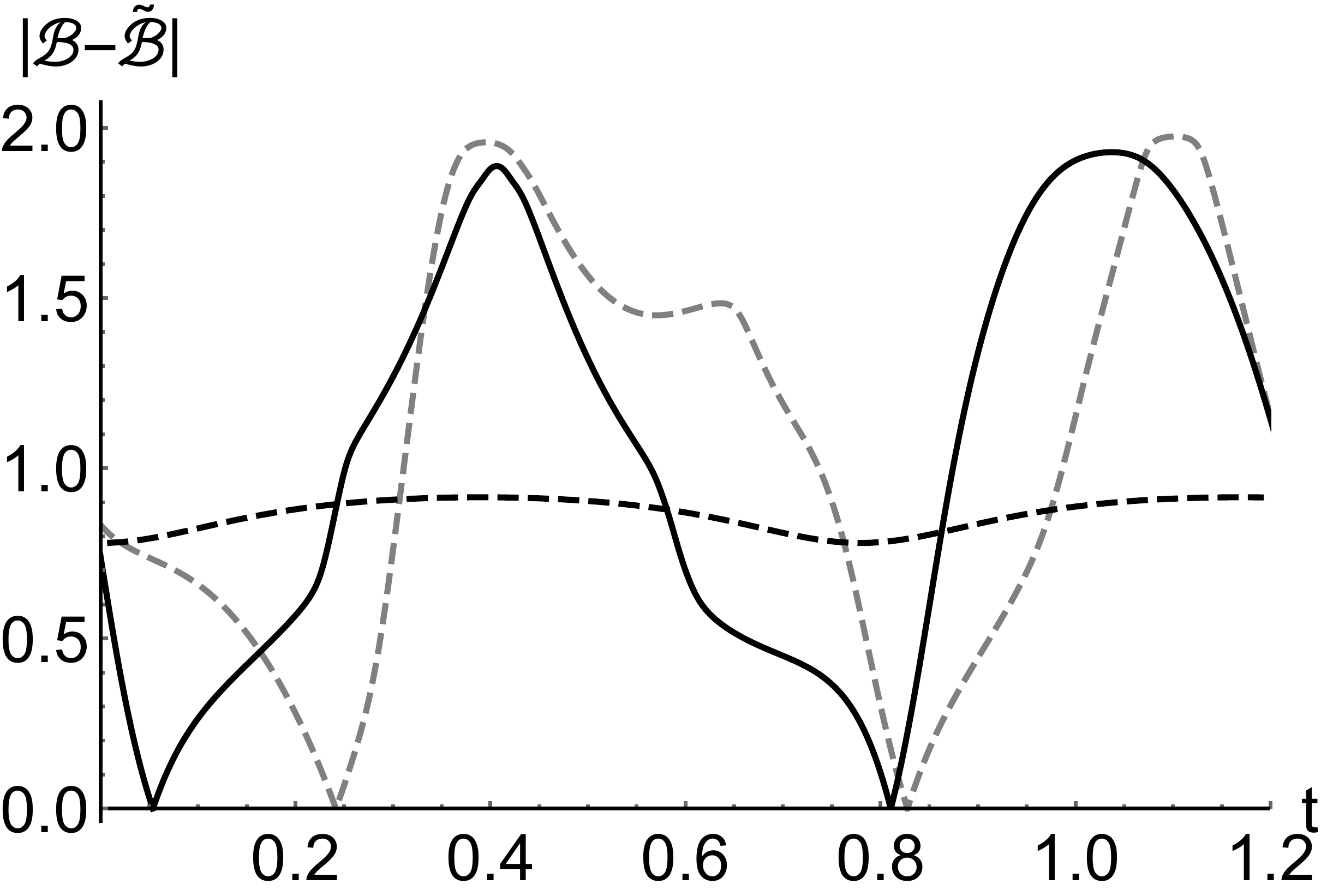} \quad
\includegraphics[scale=0.24]{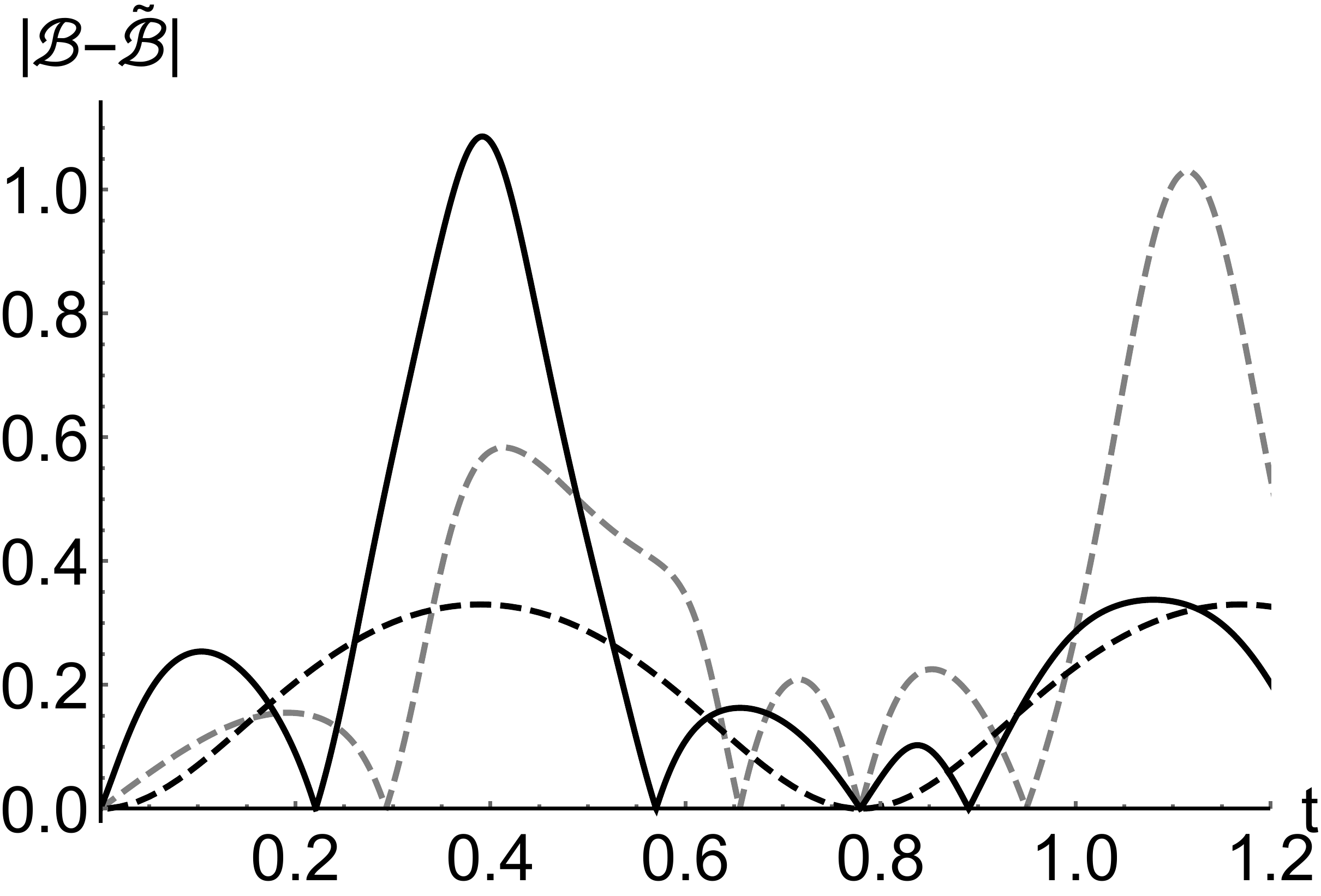}
\caption{The parameter $\vert \mathcal{B}-\widetilde{\mathcal{B}}\vert$
and the linear entropy $S_L$ (black dashed) as functions of time for the squeezed
vacuum state with squeezing parameter $\beta=4/5$~(left) and for the
coherent state with parameters $\alpha_1=4/5$ and
$\alpha_2=1/10$~(right); the amplifier parameters are: $\omega_a=1$,
$\omega_b=3$, $\Omega=9$, $k=2$, and $\nu=\sqrt{65}/2$. In both
cases, the solid black curve corresponds to a system where the
parameters $\boldsymbol{\mu}$ are given at the left of Table~\ref{tabla} and for
the gray dashed curve the parameters are given at the right of
Table~\ref{tabla}.}
\label{fig:vacio}
\end{figure}

In Fig.~\ref{fig:vacio}, the plots for $\vert
\mathcal{B}-\widetilde{\mathcal{B}} \vert$ are shown for the
evolution of the squeezed vacuum state and for the coherent state as
functions of time. It can be seen that for times equal to multiples
of the frequency $\pi / \nu$ the coherent states are separable as
indicated by the von Neumann and linear entropies and the system
state is separable for those time for the squeezed vacuum state. One
can point out that for some times the parameter $\vert
\mathcal{B}-\widetilde{\mathcal{B}} \vert$ tends to zero even when the system is
entangled (as seen in the linear and von Neumann entropies); this
behavior is present because the condition in
Eq.~(\ref{nonentangled}) is not a sufficient condition to guarantee
separability.

\subsection{No-signaling correlations}
Following the ideas of Popescu and Rohrlich that the relativistic causality does not constraint the CHSH correlations to the Cirelson bound, we are going to study the correlations of the reduced tomogram 
\[
\mathcal{W}_{1}(X_{1},\mu_{1},\nu_{1};t)=\int\mathcal{W}(X_{1},
\mu_{1},\nu_{1};X_{2},\mu_{2},\nu_{2}; t)dX_{2}\, ,
\]
without taking expression~(\ref{regions}) into account. This can be done by constructing the $4 \times 4$ stochastic matrix
$\mathbf{M}$ as follows:
\begin{equation}
\mathbf{M}=\left(\begin{array}{cccc}
P_{1}(\mu_a, \nu_b) & P_{1}(\mu_a, \nu_c) & P_{1}(\mu_d, \nu_b) & P_{1}(\mu_d, \nu_c)\\
P_{2}(\mu_a, \nu_b) & P_{2}(\mu_a, \nu_c) & P_{2}(\mu_d, \nu_b) & P_{2}(\mu_d, \nu_c)\\
P_{3}(\mu_a, \nu_b) & P_{3}(\mu_a, \nu_c) & P_{3}(\mu_d, \nu_b) & P_{3}(\mu_d, \nu_c)\\
P_{4}(\mu_a, \nu_b) & P_{4}(\mu_a, \nu_c) & P_{4}(\mu_d, \nu_b) & P_{4}(\mu_d, \nu_c)
\end{array}\right),
\label{por_red}
\end{equation}
which satisfy the normalization condition $\sum_k P_{k}(\mu_a,
\nu_b)=1$ for all column vectors and the probabilities given by
\begin{equation}
P_{i}(\mu, \nu)=\int_{\mathcal{L}_{i}}\mathcal{W}_{1}(X_{1},\mu,\nu; t)dX_{1} \ ,
\end{equation}
where $\mathcal{L}_{i}$ defines the integration region. One has that
$\sum^4_{k=1}{\mathcal{L}_{k}}= (-\infty, \infty)$ are the different regions
of the $X_1$ space where the normalization condition holds.

\begin{figure}
\centering
\includegraphics[scale=0.25]{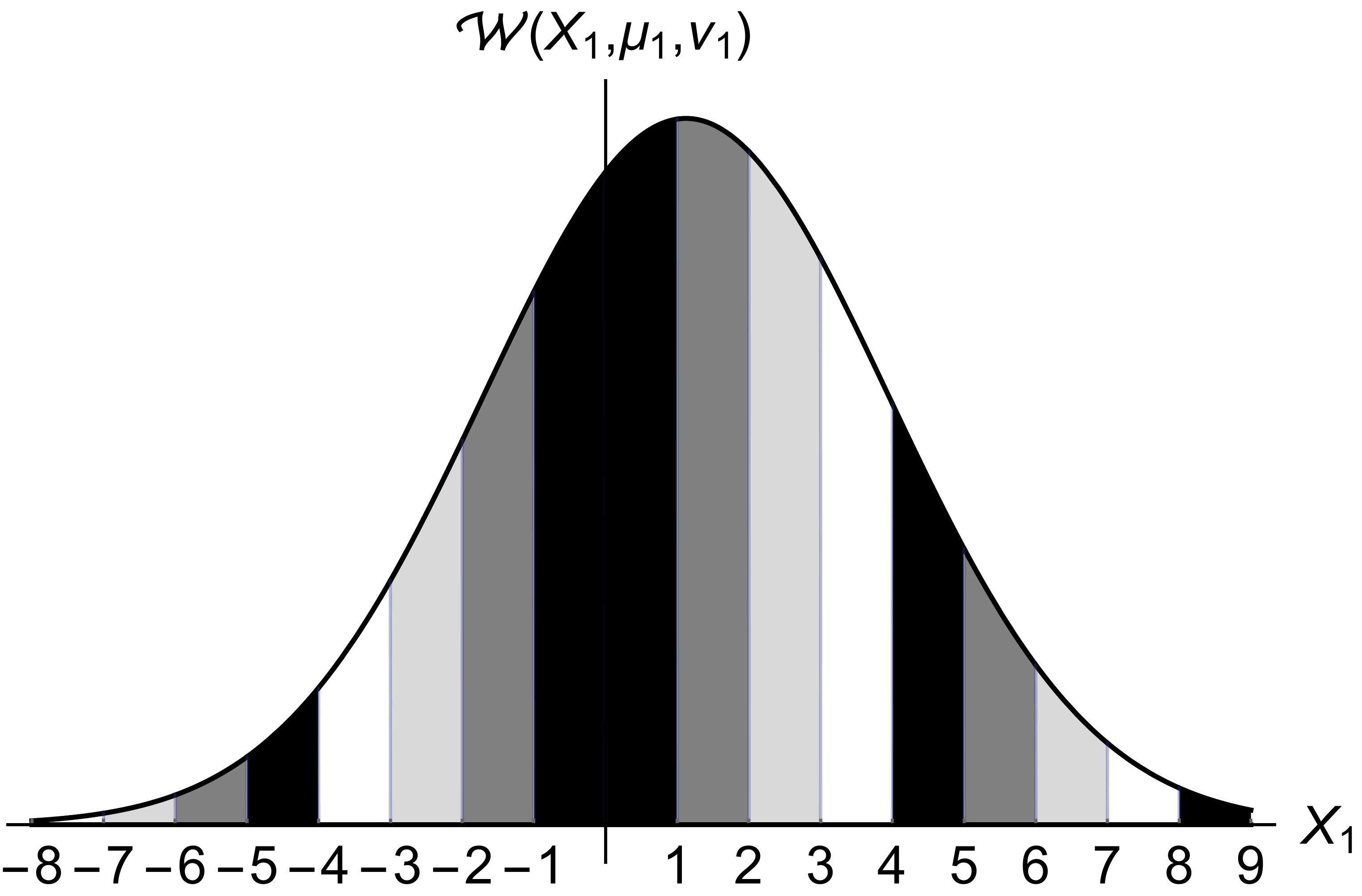} \qquad \quad
\includegraphics[scale=0.25]{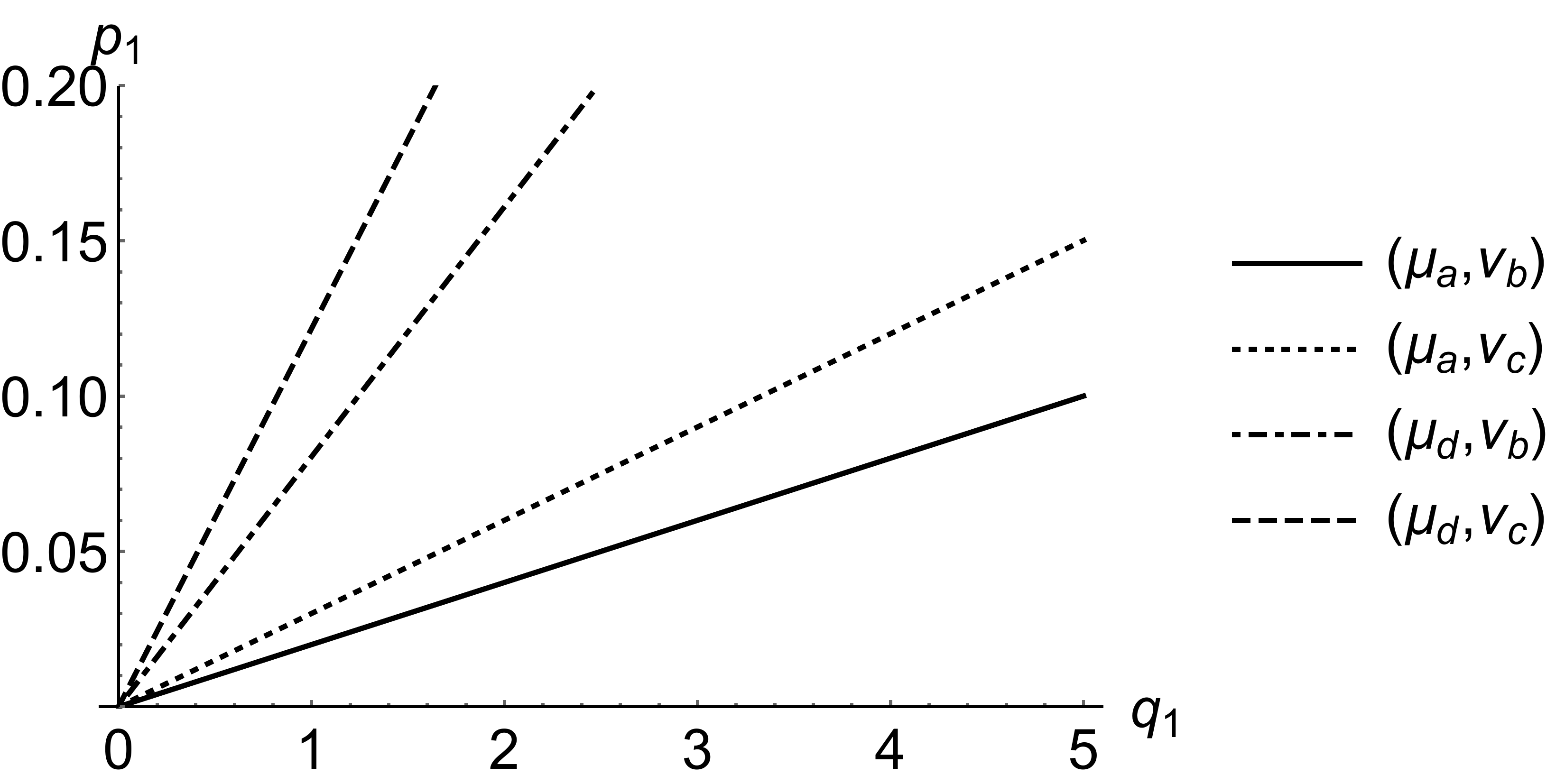}
\caption{(left) Partition used to make the portrait tomogram for the different initial states.
The region I ($\mathcal{L}_1$) is taken from 0 to $\pm$1, $\pm$4 to
$\pm$5, $\pm$8 to $\pm$9, $\cdots$ here denoted by the black area
under the reduced tomogram. The second region ($\mathcal{L}_2$) is
taken from $\pm$1 to $\pm$2, $\pm$5 to $\pm$6, $\pm$9 to $\pm$10,
$\cdots$, displayed in dark gray color. For the region III
($\mathcal{L}_3$), we have $\pm$2 to $\pm$3, $\pm$6 to $\pm$7,
$\pm$10 to $\pm$11, $\cdots$, in light gray. The last region
($\mathcal{L}_4$) is taken $\pm$3 to $\pm$4, $\pm$7 to $\pm$8,
$\pm$11 to $\pm$12, $\cdots$, which is the white region under the
tomogram.  (Right) Plots of the different $\theta$ angles defined by
the parameters $\mu$, $\nu$ in Table~\ref{tabla2}. Notice that along
these lines the tomogram is integrated in the regions mentioned
before.}
\label{fig:partition}
\end{figure}

The different $\mu$, $\nu$ parameters taken to establish the matrix
$\mathbf{M}$ of the reduced symplectic tomogram are listed in
Table~\ref{tabla2}. In Fig.~\ref{fig:partition}, the corresponding
measurement angles $\theta$ used to establish the stochastic matrix
$\mathbf{M}$ are shown. Notice that the integration variable in each
case $X_1$ is also scaled.

One can construct the Bell parameter by multiplying the matrix
$\mathbf{M}$ with $\mathbf{C}$, but now its value is only
constrained in the interval $0\leq \vert \mathcal{B} \vert \leq 4$.
Then, these correlations are not of the Bell type, given that the
partition used to construct the matrix $\mathbf{M}$ tomogram is not
of the form of a direct product of two subsystems. For that reason,
the Cirelson bound $2\sqrt{2}$ is also violated for the parameter
$\mathcal{B}$,
\begin{table}
\centering
\begin{tabular}{|c|c|c|}
\hline
($\mu,\nu$) & $\theta$ (rad) & $s$\tabularnewline
\hline
 \quad ($\mu_{a}=0.1,\ \nu_{b}=0.2$)  \quad & \quad 0.02 (1.15$^{\circ}$)  \quad &  \quad 0.1 \quad\tabularnewline
 \quad ($\mu_{a}=0.1,\ \nu_{c}=0.3$)  \quad &  \quad 0.03 (1.72$^{\circ}$)  \quad &  \quad 0.1  \quad\tabularnewline
%\hline
 \quad ($\mu_{d}=0.4,\ \nu_{b}=0.2$)  \quad & \quad 0.08 (4.58$^{\circ}$)  \quad &  \quad 0.4  \quad\tabularnewline
%\hline
 \quad ($\mu_{d}=0.4,\ \nu_{c}=0.3$)  \quad &  \quad 0.12 (6.87$^{\circ}$)  \quad &  \quad 0.4  \quad\tabularnewline
\hline
\end{tabular}
\caption{The parameters are given by $(\mu, \nu)=(s \cos\theta,s^{-1} \sin\theta)$.
The angles and scaling factors taking in the qubit portrait are indicated,
the selected values satisfy the constraint $2 \vert \mu \nu\vert \leq 1$.}
\label{tabla2}
\end{table}

% Figura 8

% Figura 9
\begin{figure}
\begin{center}
\includegraphics[scale=0.25]{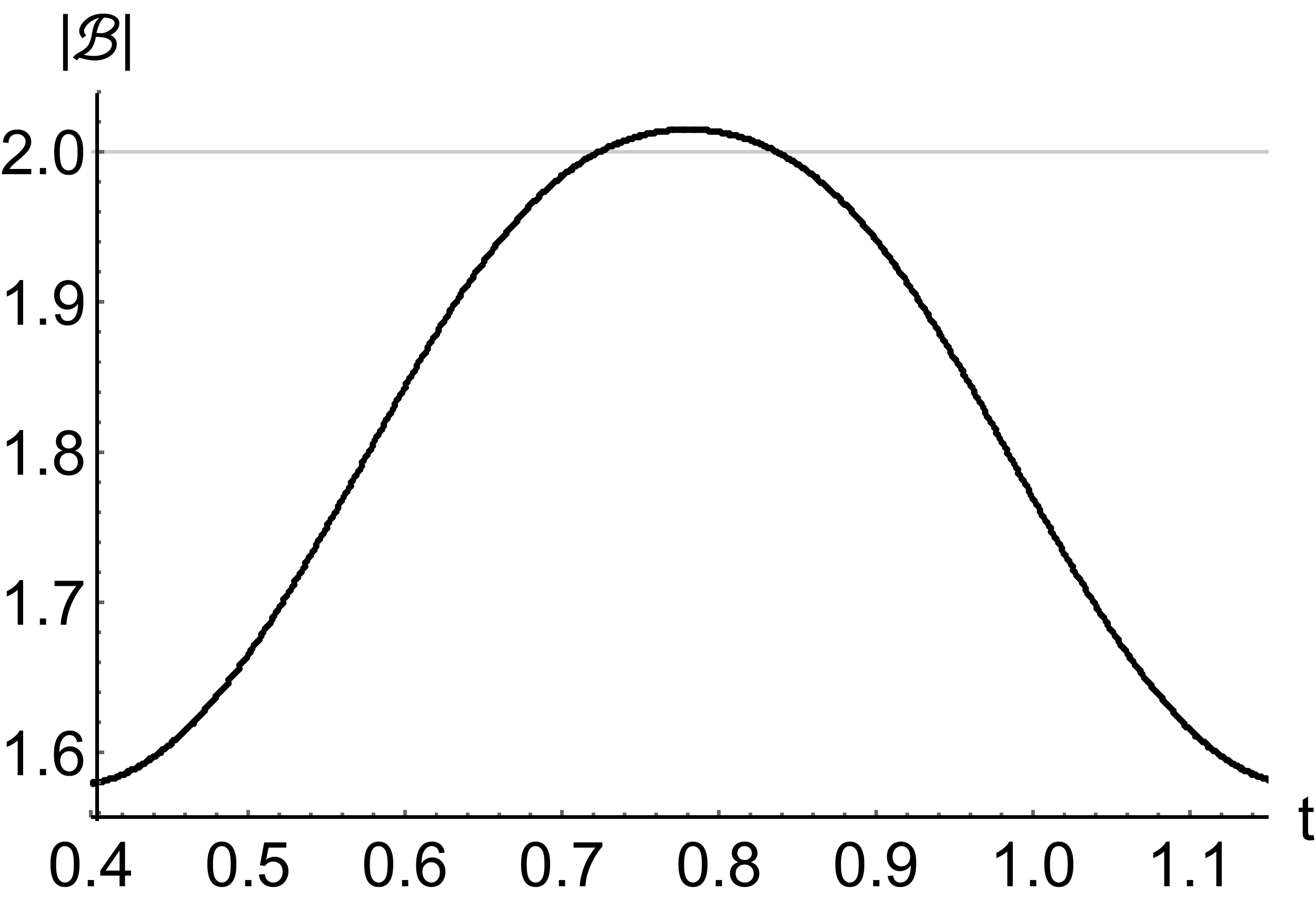} \qquad
\includegraphics[scale=0.25]{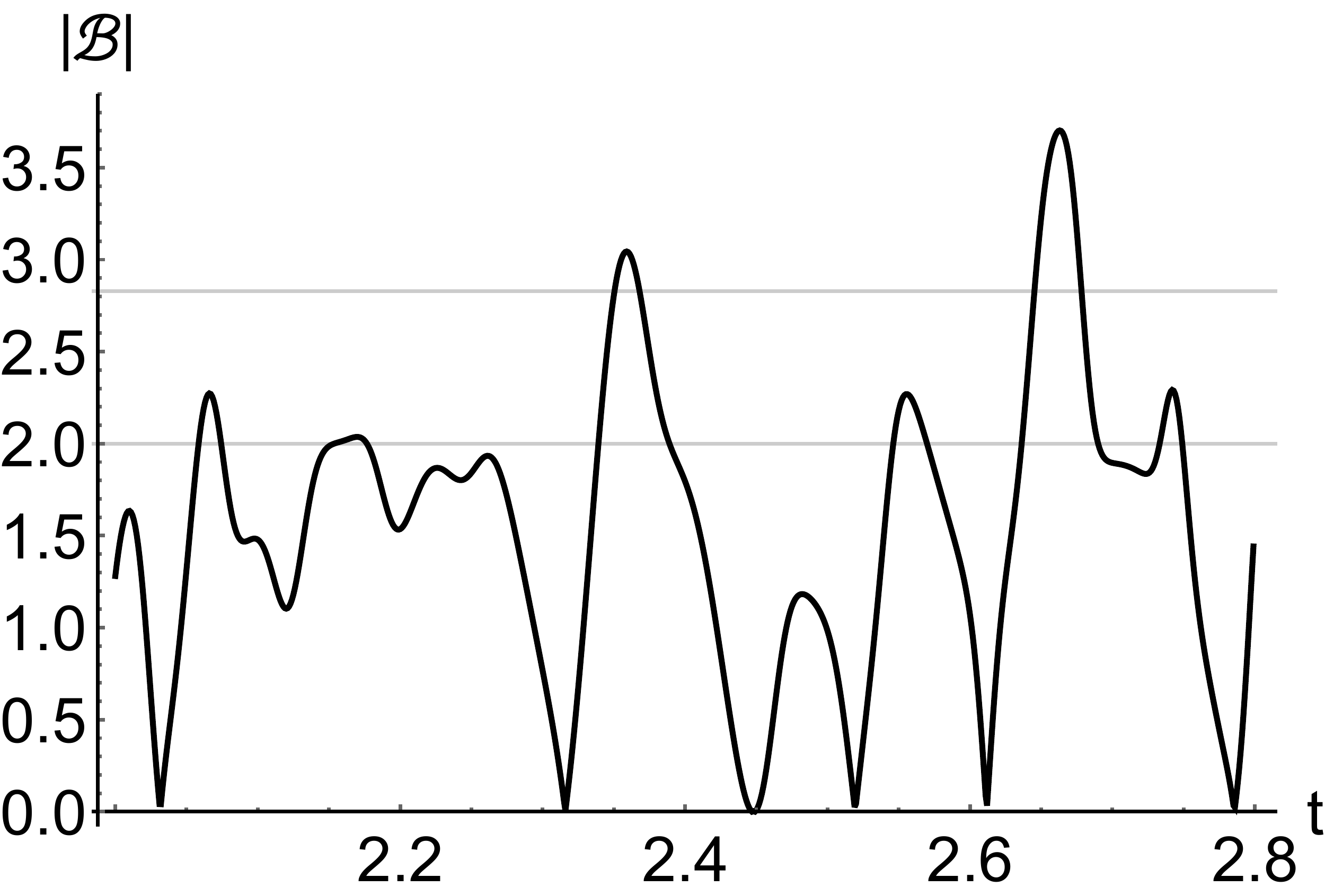}
\caption{Parameter $\mathcal{B}$ as function of time.
(Left) for the squeezed vacuum state with $\beta=4/5$ and (right)
for the coherent state with $\alpha_1=100$, $\alpha_2=3$, the
parameters of the parametric amplifier were $\Omega=1/40$, $k=1/10$,
$\omega_a=1$, and $\omega_b=3$.}
\label{fig:bellt}
\end{center}
\end{figure}
It can be seen in Fig.~\ref{fig:bellt} that there is a strong
correlation in the artificial partition of the reduced tomogram.

The behavior of the parameter $|\mathcal{B}|$ in the constructed stochastic
matrix $M$ completely corresponds to the Rohrlich--Popescu
result~\cite{popescu}. In our case, we obtained the result that the
qubit portrait of the one mode of the amplifier state tomogram is
completely different from the qubit-portrait behavior of the
qudit-system state.

\section{Conclusions}

We constructed the linear time-dependent invariants for the
non-degenerated parametric amplifier for the trigonometric and
hyperbolic cases. These invariants are used to determine in analytic
form the evolution of two-mode Gaussian wave packets, which is
always a Gaussian state characterized by the covariance
matrix in~(\ref{11}) and (\ref{17}). Also we noted that the evolution of
a squeezed vacuum state is also a squeezed vacuum state.

The corresponding tomographic representation of the states was
calculated in (\ref{30}) and (\ref{32}) and plotted in the phase space
$X_1-X_2$ displaying clearly the presence of the squeezing
phenomena.

To calculate the entanglement between the idler and signal modes of
the parametric amplifier, we establish a discretization of the
density matrix which, in principle, leads to an infinite matrix full
of zeros plus a finite $N \times N$ density matrix. By means of this
finite matrix, we have calculated the linear and von Neumann
entropies. For the evolution of the vacuum state, we have analytic
results for the entropies in Eqs. (\ref{27}) and (\ref{28}), which are compared
with the results of the discretization for two different values of
the squeezing parameter $\beta$, the mean square deviation is of the
order of $10^{-6}$.  The corresponding calculations for the coherent
and Gaussian states are also determined by the discretization
procedure. In all the cases, we have a periodic behavior (with
period $T=\pi/\nu$). If we want to determine the results for the
hyperbolic case of the parametric amplifier, we can do an analytic
continuation in the parameters of the model. We point out that the
periodic nature of the entropies is present even when the state is
not periodic.
We also have shown that the initial entanglement can be amplified due to the evolution in the parametric amplifier, the maxima values occur at times $t= (2n+1)\,\pi / (2 \nu)$ while the minima values occur at time equal to zero and $t=2n\,\pi / \nu$, $n\in\mathbb{Z}$.

In this work, we establish another procedure to determine the
entanglement of the system, which is based on a qubit portrait of a
symplectic (or optical) tomogram. This portrait uses the properties
of the stochastic matrices to define a Bell-type parameter
$\mathcal{B}$, which must satisfy the inequality $\mathcal{B} \leq
2$.  The method reduces the continuous probability of the tomogram
to a $4 \times 4$ stochastic matrix which must satisfy the previous
inequality, if it can be written as a direct product of two
stochastic matrices. This must be done by taking into account
carefully the integration regions of the continuous variables $X_1$
and $X_2$.  We have selected several possible partitions to have a
violation of the Bell-type inequality without success.

For a composite system,  the integral for joint probability
$\int_{\mathbb{A}}\mathcal{W}(1,2)\, dX_1\, dX_2$ is equal to the
product $\int_{\mathcal{L}^{(1)}}\mathcal{W}(1)\,dX_1
\int_{\mathcal{L}^{(2)}} \mathcal{W}(2)\, dX_2$. Then, one can say
that there are no correlations between the measurements of the
probabilities in the two-variable system and its state is simply
separable. Therefore, one can assume that it is simply separable and
define the Bell-type parameter $\widetilde{\mathcal{B}}$ for the
factorized tomogram, which leads to establish the equality $\vert
\mathcal{B} -
\widetilde{\mathcal{B}}\vert = 0$. This equality is a necessary
condition for the system to be separable and a violation of this
condition is a sufficient condition to determine the entanglement.

We study also the behavior of another type of correlations by
constructing a $4 \times 4$ stochastic matrix without taking into
account that the different integration regions can be written as a
direct product of two subsystems. 

As it is shown in Fig.~\ref{fig:bellt}, when the matrix $\mathbf{M}$ is
multiplied by $\mathbf{C}$ and a new parameter
$\mathcal{B}$ is defined, one can see that this pàrameter can take values larger than the Cirelson bound.

Finally, we want to enhance that the discretization procedure of the
density matrix can be expressed as a nonlinear positive mapping to
reduce even further the $N \times N$ density matrix without losing
information on the entanglement of the system as, for example, if it
has a large quantity of zeros. This method is currently explored and
it will be presented in a future publication.

\appendix

\section{Linear time-dependent invariants}

%\subsection*{}

The time-dependent invariants are operators $\Gamma$ that satisfy
\begin{equation}
\frac{d\Gamma}{dt}=0=\frac{1}{i\hbar} [\Gamma, H ]+\frac{\partial \Gamma}{\partial t}\ .
\label{a1}
\end{equation}
For quadratic Hamiltonians, $\Gamma$ is linear in the annihilation and creation operators~\cite{manko}.
Therefore, for the parametric amplifier, one proposes an invariant of the form
\[
\Gamma=\gamma_1 a+\gamma_2 b+\gamma_3 a^{\dagger}+\gamma_4 b^{\dagger} \, ,
\]
which being substituted into Eq.~(\ref{a1}) yields a coupled pair of
differential equations
\begin{eqnarray}
\dot{\gamma}_1-i\gamma_1 \omega_a-i k \gamma_4 e^{i\omega t}=0 \nonumber\, , \quad \dot{\gamma}_4+i\gamma_4 \omega_b+i k \gamma_1 e^{-i\omega t}=0 \, ,\\
\dot{\gamma}_2-i\gamma_2 \omega_b-i k \gamma_3 e^{i\omega t}=0 \nonumber\, , \quad \dot{\gamma}_3+i\gamma_3 \omega_a+i k \gamma_2 e^{-i\omega t}=0 \nonumber\, .
\label{a2}
\end{eqnarray}

To solve these sets of coupled differential equations, we consider
the transforms $\gamma_1=e^{i \omega_a t} g_1$ and $\gamma_4=e^{-i
\omega_b t} g_4$ together with $\gamma_2=e^{i
\omega_b t} g_2$ and $\gamma_3=e^{-i \omega_a t} g_3$. Substituting
these expressions into Eq.~(\ref{a2}), one arrives at
\begin{equation}
\dot{g}_1=i k e^{i \Omega t} g_4, \qquad \dot{g}_4=-i k e^{-i \Omega t} g_1,
\label{a4}
\end{equation}
where $\Omega=\omega-\omega_a-\omega_b$; and then
\[
\ddot{g}_1-i \Omega \dot{g}_1-k^2 g_1 =0
\]
whose solution, for the initial conditions $g_1 (0)=1$, $g_4 (0)=0$, is
\begin{eqnarray}
g_1 (t)= e^{i\Omega /2 t} \left(\cos \nu t-\frac{i \Omega}{2 \nu} \sin \nu t \right) \ , \nonumber \\
g_4 (t)= -\frac{i k}{\nu} e^{-i \Omega t} \sin \nu t \ . \label{a5}
\end{eqnarray}
with $\nu=\sqrt{\Omega^2 /4 - k^2}$.

Noticing that $g_2$ and $g_3$ must satisfy the same differential
equations that $g_1$ and $g_4$, respectively, but now for the
initial conditions $g_2 (0)=0$ and $g_3 (0)=0$, one gets
$g_2(t)=g_3(t)=0$.

If we now consider the initial conditions  $g_1 (0)=0$, $g_2 (0)=1$,
$g_3 (0)=0$, $g_4 (0)=0$, one has the same differential equations
making the substitutions
\[
g_2 (t)\rightarrow g_1 (t), \qquad g_3 (t)\rightarrow g_4 (t)
\]
into Eq. (\ref{a4}).

Substituting properly the previous results, one arrives at
expression~(\ref{2}) for the linear time-dependent invariants of the
parametric amplifier. The quadrature operators for the two modes are
defined as $P_1=i(A^{\dagger}-A)/\sqrt{2}$,
$P_2=i\sqrt{\frac{\omega_b}{2}}(B^{\dagger}-B)$, $Q_1=(A+A^{\dagger})/\sqrt{2}$, and
$Q_2=\frac{1}{\sqrt{2 \omega_b}}(B+B^{\dagger})$, which can be written explicitly
using Eq.~(\ref{lambdas}).

\section*{Acknowledgements}

This work was supported by CONACyT (under Project No.~238494) and
DGAPA-UNAM (under Project No.~IN110114).


\begin{thebibliography}{99}

\bibitem{trifonov}
I. A. Malkin, V. I. Man'ko, and D. A. Trifonov, Phys. Rev. D {\bf 2}
1371 (1970).

\bibitem{manko}
V. V. Dodonov and V. I. Man'ko,   {\it Invariants and the evolution
of nonstationary quantum systems}, Proceedings of the Lebedev
Physical Institute, vol. 183, (Nova Science Publishers, New York,
1989).

\bibitem{libro-manko}
V. V. Dodonov and V. I. Man'ko, (Eds.), {\it Theory of Nonclassical
States of Light} (Taylor-Francis, London, 2003).

\bibitem{suslov}
S. K. Suslov, Phys. Scr. {\bf 81} 055006 (2010).

\bibitem{castanos}
O. Casta\~nos, R. L\'opez-Pe\~na, and V. I. Man'ko, J. Phys. A: Math. Gen. {\bf 27} 1751 (1994).

\bibitem{feynman}
R. P. Feynman, A. R. Hibbs and D. F. Styer, {\it Quantum Mechanics and Path Integrals} (Dover publications, New York, 2010).

\bibitem{rekdal}
P. K. Rekdal and B. K. Skagerstam, Phys. Scr. {\bf 61} 296 (2000).

\bibitem{walls}
D. F. Walls and G. J. Milburn, {\it Quantum Optics} (Springer, Berlin, 1995).

\bibitem{zeilinger1}
 K. Takashima, N. Hatakenaka, S. Kurihara, and A.
Zeilinger, J. Phys. A: Math. Theor. {\bf 41} 164036 (2008).

\bibitem{zeilinger2}
K. Takashima, S. Matsuo, T. Fujii, N. Hatakenaka, S. Kuri- hara, and
A. Zeilinger, J. Phys.: Conf. Ser. {\bf 150} 052260 (2009).

\bibitem{zeilinger3}
T. Fujii, S. Matsuo, N. Hatakenaka, S. Kurihara, and A. Zeilinger,
Phys. Rev. B {\bf 84}  174521 (2011).

\bibitem{mancini}
S. Mancini, V. I. Man'ko, and P. Tombesi, Phys. Lett. A  {\bf 213} 1
(1996).

\bibitem{einstein}
A. Einstein, B. Podolsky, N. Rosen, Phys. Rev. {\bf 47} 777 (1935).

\bibitem{Schrod35}
E. Schr\"odinger, Naturwiss. {\bf 23} 807; 823; 844 (1935).

\bibitem{bell}
J. S. Bell, Physics {\bf 1} 195 (1964).

\bibitem{clauser}
J. F. Clauser, M. A. Horne, A. Shimony, and R. A. Holt, Phys. Rev.
Lett.  {\bf  23} 880 (1969).

\bibitem{brune} N. Brunner, D. Cavalcanti, S. Pironio, V. Scarani, S. Wehner, Rev. Mod. Phys. {\bf 86} 419 (2014).

\bibitem{aspect}
A. Aspect, P. Grangier, and G. Roger,  Phys. Rev. Lett. {\bf 47}
460 (1981).

\bibitem{guhne} O. G\"uhne, G. T\'oth, Phys. Reports {\bf 474} 1 (2009).

\bibitem{cirelson}
B. S. Cirelson, Lett. Math. Phys. {\bf 4} 93 (1980).

\bibitem{popescu}
S. Popescu and D. Rohrlich, Found. Phys. {\bf 24} 379 (1994).

\bibitem{banaszek}
K. Banaszek and K. Wodkiewicz, Phys. Rev. Lett. {\bf 82} 2009
(1999).

\bibitem{bellini}
M. D'Angelo, A. Zavatta, V. Parigi, and M. Bellini, Phys. Rev. A
{\bf 74} 052114 (2006).

\bibitem{chernega}
V. N. Chernega and V. I. Man'ko, J. Russ. Laser Res. {\bf 28} 103 (2007).

\bibitem{nielsen}
M. A. Nielsen, and I. L. Chuang, {\it Quantum Computation and Quantum Information} (Cambridge University Press, London, 2010).

\bibitem{ibort}
A. Ibort, V. I. Man'ko, G. Marmo ,  A. Simoni, and F. Ventriglia, Phys. Scr. {\bf 79} 065013 (2009).

\bibitem{filippov}
S. N. Fillipov, and V. I. Man'ko, J. Russ. Laser Res. {\bf 30} 55 (2009).

\bibitem{louisell}
W. H. Louisell, A Yariv, and A. E. Siegman, Phys. Rev. {\bf 124} 1646 (1961).

\bibitem{mollow1}
B. R. Mollow and R. J. Glauber, Phys. Rev. {\bf 160} 1076 (1967).

\bibitem{mollow2}
B. R. Mollow and R. J. Glauber, Phys. Rev. {\bf 160} 1097 (1967).

\bibitem{marhic}
M. E. Marhic, {\it Fiber optical parametric amplifiers and related devices},
 (Cambridge University Press, London, 2007).

\bibitem{marhic2}
M. Jamshidifar, A. Vedadi and M. E. Marhic, 2014 ``Continuous-Wave Two-pump Fiber Optical Parametric Amplifier with 60 dB Gain", in CLEO: 2014, OSA Technical Digest (online) (Optical Society of America, 2014), paper JW2A.21.

\bibitem{isar1}
A. Isar, Open Sys. Inf. Dynamics, {\bf 18} 175 (2011).

\bibitem{isar2}
A. Isar,  Phys. Scr. {\bf T160} 014019 (2014).

\bibitem{calo95}
O. Casta\~nos, R. L\' opez-Pe\~na and V. I. Man'ko, J. Russ. Laser Res. {\bf 16} 477 (1995).

\bibitem{julio}
O. Casta\~nos, and J. L\' opez, Phys. Conf. Ser. {\bf 380} 012017 (2012).

\bibitem{korolkova}
S. Sp\"alter, N. Korolkova, F. K\"onig, A. Sizmann, and G. Leuchs, Phys. Rev. Lett. {\bf 81} 786 (1998).

\bibitem{marino}
V. Boyer, A. M. Marino, R. C. Pooser, and P. D. Lett, Science {\bf 321} 544 (2008).

\bibitem{fabre}
F. A. S. Barbosa, A. S. Coelho, K. N. Cassemiro,  P. Nussenzveig, C. Fabre, M. Martinelli, and A. S. Villar, Phys. Rev. Lett. {\bf 111} 200402 (2013).

\bibitem{Levenson93}
J. A. Levenson, I. Abram, Th. Rivera, and Ph. Grangier, J. Opt. Soc. Am. B {\bf 10} 2233 (1993).

\bibitem{wei}
J. Wei, and E. Norman, J. Math. Phys. {\bf 4} 575 (1963).

%\bibitem{Olga1}
%V. V. Dodonov, O. V. Man'ko, and V. I. Man'ko, J. Sov. Laser Res.
%{\bf 10} 413 (1989); ibid {\bf 13} 196 (1992).
%
%\bibitem{Olga2}
%V. V. Dodonov, O. V. Man'ko, and V. I. Man'ko, Meas. Tech. {\bf 33},
%102 (1990).
%
%\bibitem{Olga3}
%V. I. Man'ko, J. Sov. Laser Res. {\bf 12} 383 (1991).
%
%\bibitem{Olga4}
%V. V. Dodonov, O. V. Man'ko, and V. I. Man'ko, Proceedings of the
%Lebedev Physical Institute, vol.~200, p.~155 (Nova Science
%Publishers, New York, 1991).
%
%\bibitem{Olga5}
%V. V. Dodonov, O. V. Man'ko, and V. I. Man'ko, Proceedings of the
%Lebedev Physical Institute, vol.~205, p.~217 (Nova Science
%Publishers, New York, 1993).
%
%\bibitem{Olga6}
%O. V. Man'ko, J. Korean Phys. Soc. {\bf 27} 1 (1994).
%
%\bibitem{Victor1}
%V. V. Dodonov, Adv. Chem. Phys. {\bf 119} 309 (2001); Phys. Scr.
%{\bf 82} 038105 (2010).
%
%\bibitem{Victor2}
%A. V. Dodonov, E. V. Dodonov, and V. V. Dodonov, Phys. Lett. A {\bf
%317} 378 (2003).
%
%\bibitem{Victor3}
%V. V. Dodonov and A. V. Dodonov, J. Russ. Laser Res. {\bf 26} 445
%(2005).
%
%\bibitem{Olga7}
%O. V. Man'ko, AIP Conf. Proc. {\bf 1424} 221 (2012); Phys. Scr. {\bf
%T153} 014046 (2013).
%
%\bibitem{merzbacher}
%E. Merzbacher, {\it  Quantum Mechanics}, (Wiley, New York, 2004).
\bibitem{vogel}
K. Vogel and H. Risken, Phys. Rev A {\bf 40} 2847 (1989).

\bibitem{dodonov}
V. V. Dodonov and V. I. Man'ko, Phys. Lett. A {\bf 239} 335 (1997).

\bibitem{castanos2}
O. Casta\~nos, R. L\'opez-Pe\~na, M. Man'ko, and V. I. Man'ko, J. Opt. B: Quantum Semiclass. Opt. {\bf 5} 227 (2003).

\bibitem{sudarshan}
V. I. Man'ko, G. Marmo, E. C. G. Sudarshan, and F. Zaccaria, Phys. Lett. A {\bf 327} 353 (2004).

\bibitem{A}
A. O. Niskanen, K. Harrabi, F. Yoshihara, Y. Nakamura, S. Lloyd, and
J. S. Tsai,
% Quantum Coherent Tunable Coupling of Superconducting Qubits
Science {\bf 316} 723 (2007).

%[DOI:10.1126/science.1141324]
\bibitem{B}
R. C. Bialczak, M. Ansmann, M. Hofheinz, M. Lenander, E. Lucero, M.
Neeley, A. D. O'Connell, D. Sank, H. Wang, M. Weides, J. Wenner, T.
Yamamoto, A. N. Cleland, and J. M. Martinis,
% Fast Tunable Coupler for Superconducting Qubits
Phys. Rev. Lett. {\bf 106} 060501 (2011).

\bibitem{C}
M. Ansmann, H. Wang, R. C. Bialczak, M. Hofheinz, E. Lucero, M.
Neeley, A. D. O'Connell, D. Sank, M. Weides, J. Wenner, A. N.
Cleland, and J. M. Martinis,
% Violation of Bellís inequality in Josephson phase qubits
Nature Lett. {\bf 461} 504 (2009).


\bibitem{drummond}
P. D. Drummond and M. D. Reid, Phys. Rev. A {\bf 41} 3930 (1990).

\bibitem{fang}
Y. Fang and J. Jing, New J. Phys. {\bf 17} 023027 (2015).

\bibitem{giedke}
G. Giedke, M. M. Wolf, O. Kr\"uger, R. F. Werner and J. I. Cirac, Phys. Rev. Lett. {\bf  91} 107901 (2003).

\bibitem{hudelist}
F. Hudelist, J. Kong, C. Liu, J. Jing, Z.Y. Ou and W. Zhang, Nature Comm. {\bf 5} 3049 (2014).

\bibitem{zhang}
J. Zhang, C. Xie and K. Peng, Phys. Lett. A, {\bf 299} 427 (2002).

\bibitem{reid}
M. D. Reid and P. D. Drummond, Phys. Rev. Lett. {\bf 60} , 2731 (1988).
%“Quantum correlations of phase in nondegenerate parametric oscillation,” 

\bibitem{cassemiro1}
K. N. Cassemiro, A. S. Villar, P. Valente, M. Martinelli and P. Nussenzveig, J. of Phys.: Conf. Ser. {\bf 84} 012003 (2007).

\bibitem{cassemiro2}
A. S. Villar, K. N. Cassemiro, K. Dechoum, A. Z. Khoury, M. Martinelli and P. Nussenzveig, J. Opt. Soc. Am. B {\bf 24} 249 (2007).

\bibitem{jing} J. Jing, J. Zhang, Y. Yan, F. Zhao, C. Xie and K. Peng, Phys. Rev. Lett. {\bf 90} 167903 (2003).

\bibitem{takei} N. Takei, H. Yonezawa, T. Aoki and A. Furusawa, Phys. Rev. Lett. {\bf 94} 220502 (2005).

\bibitem{bellini1} M. Bellini, A. S. Coelho, S. N. Filippov, V. I. Man'ko and A. Zavatta, Phys. Rev. A {\bf 85} 052129 (2012).


\end{thebibliography}
\end{document}